\documentclass[aps,prb,twocolumn,superscriptaddress]{revtex4-2}
\usepackage{bm}                     
\usepackage{graphicx}               
\usepackage{graphics}               
\usepackage{amsmath}                
\usepackage{amsfonts}               
\usepackage{amssymb}                
\usepackage{makeidx}                
\usepackage{wasysym}                
\usepackage{bbold}                  
\usepackage[makeroom]{cancel}       
\usepackage[utf8]{inputenc}         
\usepackage[normalem]{ulem}         
\usepackage{xcolor}                 
\usepackage{soul}                   
\usepackage{float}                  
\usepackage{physics}                
\usepackage{IEEEtrantools}          
\usepackage{mathrsfs}

\usepackage[T1]{fontenc}
\usepackage[utf8]{inputenc}

\newcommand{\be}{\begin{equation}}
\newcommand{\ee}{\end{equation}}
\newcommand{\ba}   {\begin{eqnarray}}
\newcommand{\ea}   {\end{eqnarray}}

\usepackage[unicode=true,pdfusetitle,
 bookmarks=true,bookmarksnumbered=false,bookmarksopen=false,
 breaklinks=false,pdfborder={0 0 1},backref=false,colorlinks=false]
 {hyperref}
\hypersetup{
 colorlinks,linkcolor=blue,citecolor=blue,urlcolor=blue}

\definecolor{navy}{RGB}{0,0,128}
\definecolor{royalblue}{RGB}{65,105,225}
\definecolor{darkorange}{RGB}{204,85,0} 
\definecolor{forestgreen}{RGB}{34,139,34}

\begin{document}

\preprint{APS/123-QED}
\title{
Quadrupole Origin of the Orbital Hall Effect in an sp$^3$ Centrosymmetric Lattice: Robustness and Tunability under Disorder
}

\author{Kevin J. U. Vidarte}
\email{kevin.urcia@inl.int}
\affiliation{International Iberian Nanotechnology Laboratory (INL), Av. Mestre José Veiga, 4715-330 Braga, Portugal}
\author{Henrique P. Veiga}
\affiliation{Centro de Física das Universidades do Minho e Porto
Departamento de Física e Astronomia, Faculdade de Ciências,
Universidade do Porto, 4169-007 Porto, Portugal}
\affiliation{School of Physics, Engineering and Technology and York Centre for Quantum Technologies, University of York, York YO10 5DD, United Kingdom}
\author{Aur\'elien Manchon}
\affiliation{Aix-Marseille Universit\'e, CNRS, CINaM, Marseille, France}
\author{Tatiana G. Rappoport}
\email{tatiana.rappoport@inl.int}
\affiliation{International Iberian Nanotechnology Laboratory (INL), Av. Mestre José Veiga, 4715-330 Braga, Portugal}
\affiliation{Centro Brasileiro de Pesquisas Físicas (CBPF), Rua Dr Xavier Sigaud 150, Urca, 22290-180, Rio de Janeiro-RJ, Brazil}
\affiliation{Physics Center of Minho and Porto Universities (CF-UM-UP),Campus of Gualtar, 4710-057, Braga, Portugal}

\date{\today}
\begin{abstract}
We investigate how disorder affects orbital Hall transport in the centrosymmetric spinless \(sp^3\) tight-binding model on a square lattice, widely used as a minimal model for the orbital Hall effect~\cite{Go2018}. In the pristine system, the intrinsic orbital Hall response is related to the quadrupolar orbital texture generated by \(s\)-\(p\) hybridization. Using the Kubo-Bastin formalism expanded in Chebyshev polynomials, we analyze the effect of vacancies, orbital-dependent Anderson on-site disorder, and selective \(V_{sp\sigma}\)-bond disorder on the orbital transport. For Anderson disorder and vacancies, the longitudinal charge conductivity decreases faster than the orbital Hall conductivity over the range studied, leading to an increase of the orbital Hall angle and, in the Anderson case, saturation at stronger disorder.
Selective \(V_{sp\sigma}\)-bond disorder behaves differently: by modifying only the \(s\)-\(p\) hybridization channel while leaving the remaining hopping network intact, it can produce a non-monotonic orbital Hall response and enhance the orbital Hall conductivity at intermediate dilution. Overall, our results show that the orbital Hall response remains sizable under vacancy and on-site disorder, while selective modification of the \(s\)-\(p\) hybridization provides an additional route to tune the orbital Hall conductivity.
\end{abstract}



\maketitle




\section{Introduction}
\label{Sec:introduction}

Orbitronics studies non-equilibrium phenomena associated with electronic orbital angular momentum in solids, including its generation, transport, accumulation, and detection~\cite{GoReview2021,OrbitronicsPerspective2025,Cysne2025}.
A widely studied transport response is the orbital Hall effect (OHE), in which an applied electric field drives a transverse flow of orbital angular momentum~\cite{Bernevig2005,Kontani2009,Go2018,Choi2023,Lyalin2023,Sala2023}.
Closely related current-induced orbital responses include the orbital
Rashba--Edelstein effect~\cite{Chen2018,Salemi2019,Ding2022,ElHamdi2023,Gao2025}
and orbital torque~\cite{GoLee2020,Ding2020,Lee2021}.

Unlike its spin counterpart, the OHE does not require strong spin-orbit
coupling, so it can be large in light elements and in materials where
spin-orbit-driven responses are weak~\cite{Jo2018,Rappoport2023}. However, those same materials are rarely the clean crystals in which the
intrinsic mechanism is defined: the metals in which the OHE is measured and
used are alloyed, polycrystalline, or amorphous. Whether the intrinsic response survives in such systems, and how it is modified by disorder, is therefore an important question. 

Recent theoretical works have addressed different disorder regimes, including the robustness of intrinsic orbital Hall contributions in weakly disordered crystalline metals, the role of vertex corrections, and the emergence of extrinsic orbital Hall mechanisms in dilute or ultraclean limits~\cite{Tang2024,Mankovsky2024,Liu2024,Liu2024b,Veneri2025,Barbosa2025,Joy2026}. 
At the same time, experiments are now probing orbital Hall signals in systems where the conductivity is tuned by alloying, structural disorder, or crystalline order~\cite{Peng2025,Yang2026,Kang2026}. 
These results show that orbital Hall responses can persist in disordered samples, but also that their scaling with longitudinal transport depends on the microscopic disorder and transport regime.

The intrinsic OHE has been connected to orbital textures in momentum space, understood as the momentum-dependent orbital composition of Bloch states~\cite{Go2018}. 
In the \(p\)-orbital sector, this texture can be decomposed into dipolar and quadrupolar components.  For centrosymmetric and time-reversal-symmetric systems, the equilibrium orbital dipole is quenched, so the OHE cannot be understood as arising from a pre-existing texture of \(L_z\). The quadrupoles provide the relevant structure: quadrupole expectation values can form non-trivial momentum-space textures without breaking either symmetry. 
Since the dynamics of the orbital dipole is coupled to the quadrupole degrees of freedom~\cite{Han2022,Manchon2026}, these quadrupolar textures provide the microscopic structure from which the transverse \(L_z\) orbital current emerges.
Very recently, polarization-resolved Kerr microscopy has directly observed current-induced orbital-quadrupole accumulation in Ti and Pt, isolating the time-reversal-even quadrupolar response from the dipolar contribution and finding the quadrupolar optical response larger in both metals~\cite{Lee2026Quadrupole}.

Here we make this connection explicit in a spinless \(sp^3\) tight-binding model on a square lattice. The \(s\)-\(p\) hybridization generates a quadrupolar texture and a low-energy theory relates this texture to the intrinsic \(L_z\)-projected orbital Hall response. This clean limit provides the reference point for studying how different microscopic forms of disorder modify orbital Hall transport.

We consider orbital-dependent Anderson disorder, vacancies, and selective \(V_{sp\sigma}\)-bond disorder. The first two provide standard forms of static disorder, while the last one acts directly on the \(s\)-\(p\) hybridization channel. Using the Kubo--Bastin formalism expanded in Chebyshev polynomials, we evaluate the orbital Hall conductivity, the longitudinal conductivity, and the orbital Hall angle, allowing us to distinguish changes in orbital-current generation from changes in charge transport as disorder is increased. We find that the orbital Hall conductivity is markedly less sensitive to disorder
than the longitudinal conductivity, so that the orbital Hall angle increases over a wide disorder range before saturating, consistent with trends reported in recent
experiments~\cite{Yang2026,Kang2026}.
\section{Theory and Method}
\label{Sec:theory&model}

    \subsection{Model Hamiltonian}
    \label{SubSec:model_Hamiltonian}
    To analyze the orbital Hall response and its relation to orbital textures, we consider a tight-binding model on a simple square lattice with four orbitals $s, p_x, p_y, p_z$ at each lattice point.
    We ignore the spin degree of freedom to focus solely on orbital response.
    The real-space Hamiltonian reads
    \begin{align}
        \label{eq:real-sapce_hamiltonian}
        H=\sum_{i,\alpha} \varepsilon^{\alpha}_{i} c^{\dagger}_{i,\alpha} c_{i\alpha} 
        -\!\!\!\!\sum_{\langle i,j\rangle,\alpha,\beta}\left[ t^{\alpha\beta}_{ij} c^{\dagger}_{i\alpha} c_{j\beta} + {\rm h.c.}\right]
        ,
    \end{align}
    where $\varepsilon^{\alpha}_{i}$ represents the on-site energy of the $\alpha$ orbital at site $i$, $t^{\alpha\beta}_{ij}$ is the transfer integral from orbital $\beta$ at site $j$ to orbital $\alpha$ at site $i$.
    To determine the hopping amplitudes we adopt the Slater-Koster approximation \cite{SlaterKoster1954},
    \begin{align}
        \label{eq:slater-koster_parameters}
        &t_{ij}^{ss} = V_{ss\sigma}, \nonumber \\
        &t_{ij}^{sp_x} = l V_{sp\sigma}, \nonumber \\
        &t_{ij}^{p_xp_x} = l^2 V_{pp\sigma} + (1 - l^2)V_{pp\pi}, \nonumber \\
        &t_{ij}^{p_xp_y} = lm (V_{pp\sigma} - V_{pp\pi}), 
    \end{align}
    where $l$ and $m$ are the direction cosines of the vector connecting sites $i$ and $j$. 
    Numerical values of the Hamiltonian parameters are as follows: $\varepsilon_i^s=3.2$~eV, and $\varepsilon_i^{p_x}=\varepsilon_i^{p_y}=\varepsilon_i^{p_z}=-0.5$~eV for on-site energies, $V_{ss\sigma}=-0.5$~eV, $V_{sp\sigma}=0.5$~eV, $V_{pp\sigma}=0.5$~eV, and $V_{pp\pi}=-0.2$~eV for nearest-neighbor hopping amplitudes.
    
    When we write this Hamiltonian in \(\mathbf{k}\)-space, the different parities of the \(s\) and \(p\) orbitals appear as complex electronic couplings.
The \(s\) orbital has even parity, while the \(p_x\) and \(p_y\) orbitals have odd parity, so the \(s\)-\(p\) hopping changes sign when the bond direction is reversed.
For a bond along the direction \(\mu=x,y\), the two opposite contributions are \(+V_{sp\sigma}e^{ik_\mu a}\) and \(-V_{sp\sigma}e^{-ik_\mu a}\), whose sum gives
\(
H_{sp_\mu}(\mathbf{k}) = 2iV_{sp\sigma}\sin(k_\mu a) \) with \(\mu=x,y\).
The \(s\)-\(p\) mixing is therefore purely imaginary and momentum dependent.
It acts as an effective momentum-space vector potential and generates a finite orbital Berry curvature in the model~\cite{Go2018}.

    \subsection{Microscopic features of the quadrupole texture}
    \label{SubSec:Microscopic_features}

    While related works have discussed the OHE in terms of orbital textures~\cite{Han2023}, here we decompose these textures into multipolar components \cite{Manchon2026}. In a full \(p\)-orbital manifold, the non-charge orbital degrees of freedom can be organized into three orbital dipoles and five orbital quadrupoles: population imbalances among \(p_x\), \(p_y\), and \(p_z\) are described by diagonal quadrupoles such as \(Q_{x^2-y^2}\) and \(Q_{3z^2-r^2}\), real coherences between pairs of orbitals by \(Q_{xy}\), \(Q_{yz}\), and \(Q_{zx}\), and chiral complex combinations by the components of \(\mathbf{L}\). Thus, real linear combinations of \(p\) orbitals correspond to quadrupolar textures, while chiral complex combinations carry orbital angular momentum. These operators are equivalent to the orbital angular position (OAP) operators introduced in Ref.~\cite{Han2022}.
    This multipolar decomposition is useful because $Q_{xy}$ and $Q_{x^2-y^2}$ are Hermitian orbital operators with a direct physical meaning: they describe, respectively, real inter-orbital coherence and orbital population imbalance. In this sense, the quadrupole language provides a concrete operator description of the orbital texture rather than only a characterization of the Bloch-state composition.

    While the orbital dipole vanishes at equilibrium in the presence of both time-reversal and inversion symmetries~\cite{Go2018,Han2022}, the momentum-dependent \(s\)-\(p\) hybridization generates a non-uniform quadrupolar texture. 
    The symmetry argument behind this distinction is simple. 
    Under inversion the \(p\) orbitals are odd, so any bilinear combination in them is even, and both
    \(\mathbf{L}\) and \(\mathbf{Q}\) are inversion even; inversion symmetry alone does not force either response to vanish. Under time reversal, however, \(\mathbf{L}\) is odd while
    \(\mathbf{Q}\) is even. Combining the two,
    inversion symmetry requires \(\langle L_z\rangle_{\mathbf{k}} =
    \langle L_z\rangle_{-\mathbf{k}}\) while time-reversal symmetry requires
    \(\langle L_z\rangle_{\mathbf{k}} = -\langle L_z\rangle_{-\mathbf{k}}\), so the
    orbital dipole vanishes at every \(\mathbf{k}\). For the quadrupole both
    symmetries impose the same condition, \(\langle Q\rangle_{\mathbf{k}} =
    \langle Q\rangle_{-\mathbf{k}}\), which is satisfied by any even function of
    \(\mathbf{k}\) and leaves the quadrupole texture unconstrained.

    We first analyze the clean system in momentum space by Fourier transforming the
real-space lattice model of Eq.~\eqref{eq:real-sapce_hamiltonian} and obtaining the Bloch Hamiltonian \(H(\mathbf{k})\).
    Using the basis $\Psi_\mathbf{k} = (|s\mathbf{k}\rangle, |p_x\mathbf{k}\rangle, |p_y\mathbf{k}\rangle, |p_z\mathbf{k}\rangle)^T$, the matrix elements are given by the Slater-Koster parameters specified in Eq.~\eqref{eq:slater-koster_parameters}. 

    Because the model is two-dimensional, the \(p_z\) orbital remains decoupled
from the \(s,p_x,p_y\) sector. As a result, the only quadrupole operators that can develop
non-trivial momentum-dependent expectation values in the \(p_x,p_y\)
subspace are 
\(
Q_{xy} = \{L_x,L_y\}
= -\hbar^2\left(|p_x\rangle\langle p_y|
+ |p_y\rangle\langle p_x|\right),
\)
which describes the real linear combination between \(p_x\) and \(p_y\), and
\(
Q_{x^2-y^2}=L_x^2-L_y^2
= -\hbar^2\left(|p_x\rangle\langle p_x|
-|p_y\rangle\langle p_y|\right),
\)
which describes the imbalance between \(p_x\) and \(p_y\).
The remaining quadrupoles either involve \(p_z\) or reduce to a trivial
constant in this subspace, and therefore do not generate additional textures
for the present model.

    \begin{figure}[t]
    \centering
    \includegraphics[width=1.0\linewidth]{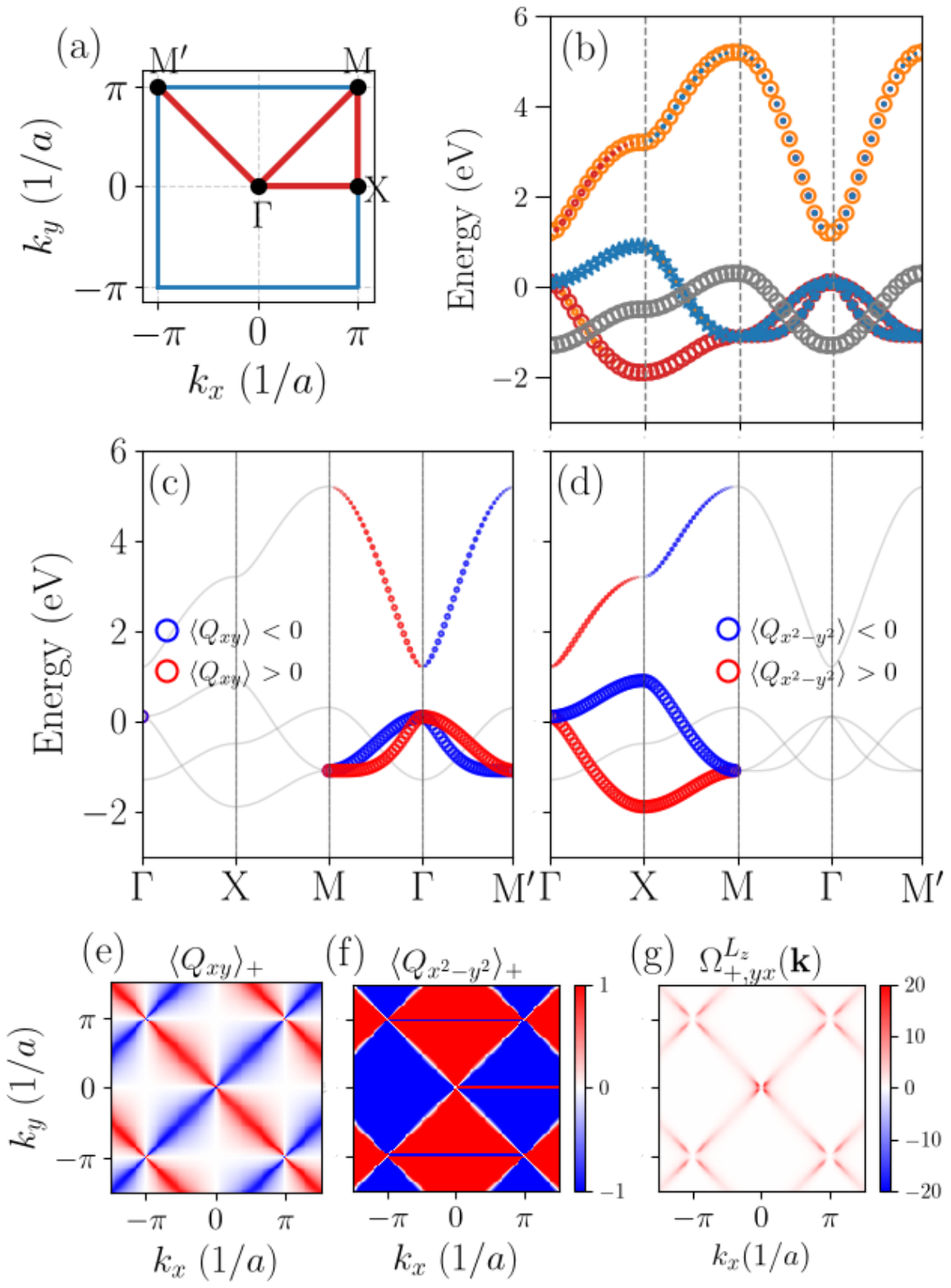}
    \caption{
    Clean-limit band structure, orbital quadrupole texture, and \(L_z\)-projected Berry curvature of the full lattice model.
    (a) Square-lattice Brillouin zone showing the $\Gamma$-${\rm X}$-${\rm M}$-$\Gamma$-${\rm M}'$ path.
    (b) Band structure along the path, marker color indicating dominant orbital character ($s$: orange; $p_x$: red; $p_y$: blue star; $p_z$: gray) and marker size the corresponding orbital weight at each $(\mathbf{k},\varepsilon)$.
    (c),(d) $\langle Q_{xy}\rangle$ and $\langle Q_{x^2-y^2}\rangle$ along the same path, color indicating sign (blue negative, red positive), marker size the orbital weight.
    (e)-(g) $\langle Q_{xy}\rangle_+$, $\langle Q_{x^2-y^2}\rangle_+$, and the $L_z$-projected Berry curvature of the upper band, $\Omega^{L_z}_{+,yx}(\mathbf{k})$, evaluated by exact diagonalization over the entire Brillouin zone.
    $\Omega^{L_z}_{+,yx}(\mathbf{k})$ is evaluated via the exact Kubo sum.
    }
    \label{fig:Fig_00}
    \end{figure}
    
Figure~1 summarizes the clean-limit structure of the full lattice model. 
Panel~1(a) defines the high-symmetry path used throughout the paper, while Fig.~1(b) shows the corresponding band structure resolved by orbital character. 
The \(p_x\) and \(p_y\) bands are degenerate at both \(\Gamma\) and \(M\), where \(\sin(k_xa)=\sin(k_ya)=0\) and the \(s\) orbital decouples from the \(p\)-orbital sector. 
Along the diagonal \(M-\Gamma-M'\), the condition \(H_{xx}(\mathbf{k})=H_{yy}(\mathbf{k})\) follows from the \(x\leftrightarrow y\) symmetry of the \(p_x,p_y\) block. 
Away from these symmetry-enforced degeneracies, the splitting of the \(p_x,p_y\) bands is generated by the \(s\)-\(p\) hybridization.

Figures~1(c) and 1(d) show the corresponding quadrupole expectation values along the same path. 
The two bands carry opposite quadrupolar textures: \(Q_{xy}\) is selected along the diagonal directions, while \(Q_{x^2-y^2}\) distinguishes the \(p_x\)- and \(p_y\)-dominated sectors along the principal directions. 
The full Brillouin-zone maps in Figs.~1(e) and 1(f) make this texture more explicit: \(Q_{xy}\) changes sign between neighboring quadrants, whereas \(Q_{x^2-y^2}\) changes sign across the zone diagonals. 
Figure~1(g) shows that the \(L_z\)-projected Berry curvature is concentrated in the same regions where the quadrupolar texture varies most strongly. 
Thus, the full lattice calculation already displays the main microscopic ingredients of the intrinsic OHE: \(s\)-\(p\) hybridization produces a quadrupolar texture in the \(p_x,p_y\) sector, and the orbital Berry curvature follows from this texture. Appendix~\ref{App:kite-arpes} describes how these textures are extracted from momentum- and orbital-resolved spectral functions using Quantum KITE, while Appendix~\ref{App:disorder-textures} applies the same analysis to the disordered systems.
    
    The connection between $sp$ hybridization and these quadrupole degrees of freedom becomes explicit when examining the low-energy effective Hamiltonian near the center of the Brillouin zone ($\Gamma$ point). 
    By employing the Löwdin downfolding technique \cite{Han2023} to project out the higher-lying $s$ band as a remote mediator, the effective Hamiltonian projected onto the $p$-subspace is, to second order,
    \begin{equation}
        \label{eq:effective_hamiltonian}
        H_p^{\rm eff}(\mathbf{k}) = h_0(\mathbf{k})\,\mathbb{1} 
        - \frac{C}{\hbar^2}k_xk_y\,Q_{xy} 
        - \frac{\mathcal{B}}{\hbar^2}(k_x^2-k_y^2)\,Q_{x^2-y^2},
    \end{equation}
    with $h_0(\mathbf{k})=E_p+\mathcal{A}(k_x^2+k_y^2)$, $\mathcal{A}=A+C/2$, $\mathcal{B}=B+C/2$, $C=4a^2V_{sp\sigma}^2/(\varepsilon_p-\varepsilon_s)$, $A=\tfrac{a^2}{2}(V_{pp\sigma}+V_{pp\pi})$, $B=\tfrac{a^2}{2}(V_{pp\sigma}-V_{pp\pi})$, and $E_p=\varepsilon_p + 2(V_{pp\sigma}+V_{pp\pi})$.
    Thus the projected Hamiltonian has a simple Pauli-matrix structure in the $\{|p_x\rangle,|p_y\rangle\}$ subspace, where $Q_{xy}=-\hbar^2\sigma_x$ and $Q_{x^2-y^2}=-\hbar^2\sigma_z$.
    
\begin{figure}[t]
    \centering
    \includegraphics[width=1.0\linewidth]{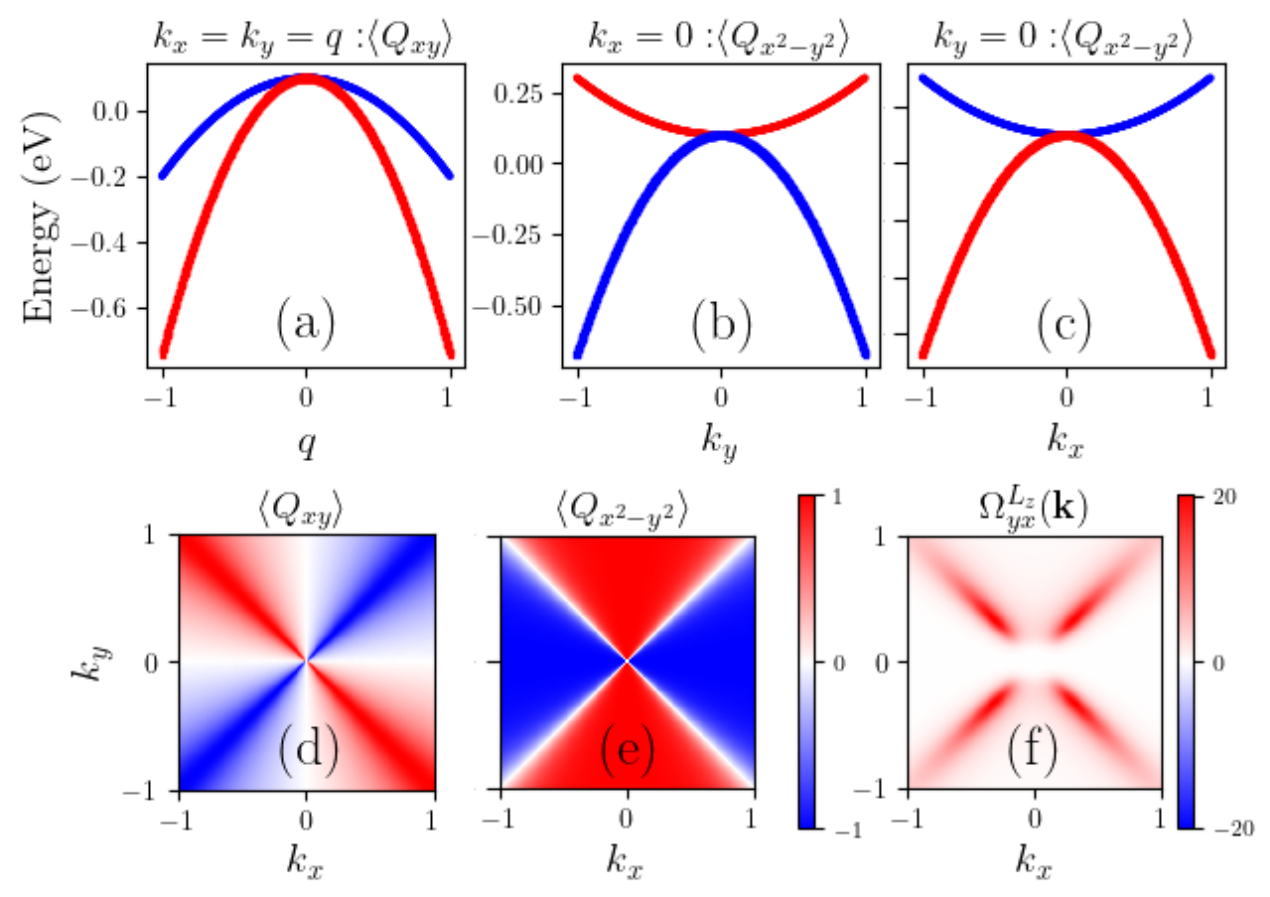}
    \caption{
    Band structure and band-resolved quadrupolar textures of the effective $p_x$-$p_y$ model, valid near the $\Gamma$ point, along {(a) $k_x=k_y$, (b) $k_x=0$ and (c) $k_y=0$.
    (a) corresponds to $\langle Q_{xy}\rangle$ and (b)-(c) correspond to $\langle Q_{x^2-y^2}\rangle$.}
    The two bands carry opposite textures, red positive and blue negative.
    {(d), (e), and (f) show the second-band quadrupole and Berry-curvature textures from the effective theory, Eqs.~\eqref{eq:textures} and \eqref{eq:omega_final} -- $\langle Q_{xy}\rangle$, $\langle Q_{x^2-y^2}\rangle$, and $\Omega^{L_z}_{yx}(\mathbf{k})$, respectively -- restricted to the small-$\mathbf{k}$ region $k_x,k_y\in[-1,1]$ near $\Gamma$ where this quadratic expansion is valid, rather than the full Brillouin zone shown in Fig.~\ref{fig:Fig_00}.}
    }
    \label{fig:Fig_01}
    \end{figure}

    Diagonalizing Eq.~\eqref{eq:effective_hamiltonian} gives the exact bands $E_\pm(\mathbf{k})=h_0(\mathbf{k})\pm d(\mathbf{k})$, $d(\mathbf{k})=\sqrt{d_x^2+d_z^2}$, with $d_x(\mathbf{k})=Ck_xk_y$ and $d_z(\mathbf{k})=\mathcal{B}(k_x^2-k_y^2)$. 
    The eigenstates carry quadrupole expectation values
    \begin{equation}
        {\langle Q_{xy}\rangle_\pm = \mp\hbar^2\frac{d_x}{d}, \qquad
        \langle Q_{x^2-y^2}\rangle_\pm = \mp\hbar^2\frac{d_z}{d}.}
        \label{eq:textures}
    \end{equation}
    Figure~\ref{fig:Fig_01} shows the resulting textures from this effective theory, restricted to the small-$\mathbf{k}$ window where the quadratic expansion holds.

Near \(\Gamma\), where the quadratic expansion is valid, the effective theory agrees with the exact diagonalization of the full lattice model. The band-resolved quadrupole textures in Figs.~\ref{fig:Fig_01}(a)-(c) reproduce the corresponding features of Figs.~\ref{fig:Fig_00}(c),(d), while the upper-band maps in Figs.~\ref{fig:Fig_01}(d)-(f) reproduce the local structure of Figs.~\ref{fig:Fig_00}(e)-(g). 
    
Figure~\ref{fig:Fig_01}(a)-(c) shows the band-resolved quadrupole textures of the effective model near \(\Gamma\). 
Along the principal axes, the texture is dominated by \(Q_{x^2-y^2}\), while along the diagonal \(k_x=k_y\) it is dominated by \(Q_{xy}\). 
Away from these high-symmetry directions, both quadrupole components are present, and the two bands carry opposite textures. 
The corresponding momentum-space maps in Figs.~\ref{fig:Fig_01}(d) and (e) show the expected fourfold pattern: \(\langle Q_{xy}\rangle\) changes sign between neighboring quadrants, whereas \(\langle Q_{x^2-y^2}\rangle\) changes sign across the diagonals. 
The angular profile is slightly distorted because the coefficients of the two quadrupole terms are different, reflecting the competition between direct \(p\)-orbital hopping and the virtual \(p\)-\(s\)-\(p\) process.

    This quadrupole structure is the microscopic origin of the orbital Hall response. 
    The velocity operators $v_a=\hbar^{-1}\partial_{k_a}H_p^{\rm eff}$ inherit the momentum derivatives of $d_x$ and $d_z$, so the electric-field vertex $v_x$ directly probes the quadrupole texture. 
    The orbital-current operator $j_y^{L_z}=\tfrac{1}{2}\{v_y,L_z\}$, by contrast, anticommutes with both quadrupole operators and collapses onto the orbital-dipole channel alone, $j_y^{L_z}=(2\mathcal{A}k_y/\hbar)L_z$: the current vertex identifies what is transported ($L_z$) but carries no memory of the texture itself. 
    The texture re-enters through the interband matrix element $\langle u_-|v_x|u_+\rangle=(d_z\partial_{k_x}d_x-d_x\partial_{k_x}d_z)/(\hbar d)$, finite only when the field rotates the quadrupole eigenbasis, and vanishing identically if $C=0$ or $\mathcal{B}=0$. 
    Combining both vertices gives the projected Berry curvature
    \begin{equation}
        {\Omega^{L_z}_{\pm,yx}(\mathbf{k}) = \mp\frac{\mathcal{A}C\mathcal{B}\,k_y^2(k_x^2+k_y^2)}{d(\mathbf{k})^3},}
        \label{eq:omega_final}
    \end{equation}
    shown in Fig.~\ref{fig:Fig_01}(f) with weight concentrated away from $\Gamma$ and the axes, where the interband coupling requires both quadrupole channels to be simultaneously active.

This clean, disorder-free texture provides the microscopic reference for the disorder analysis below. 
Equation~\eqref{eq:omega_final} shows that, in the low-energy clean limit, the \(L_z\)-projected Berry curvature depends on both the \(s\)-\(p\)-mediated coefficient \(C\) and the \(p_x\)-\(p_y\) quadrupolar coefficient \(\mathcal{B}\) and disorder can affect these ingredients in different ways.

    \subsection{The Kubo-Bastin formula for Orbital-Conductivity}
    \label{SubSec:Kubo-Bastin_formula}
    To evaluate the orbital conductivity $\sigma_{xy}^{\mathcal{O}}$ in large-scale disordered systems, we employ the Kubo-Bastin linear response formalism.
    The orbital current operator flowing along the $\alpha$-direction with orbital operator $\mathcal{O}$ is defined as
    \begin{align}
        J_{\alpha}^{\mathcal{O}}=\dfrac{1}{2}\{\mathcal{O},v_\alpha\}
    \end{align}
    where $v_\alpha = \frac{i}{\hbar}\left[ H, r_\alpha\right]$ is the velocity operator component.
    Within the independent-particle approximation, the dc orbital conductivity is given by Bastin formula \cite{Bastin1971},    
    \begin{align}
        &\sigma_{\alpha\beta}^{\mathcal{O}} (\mu ,T) = \dfrac{ie\hbar}{\Omega} \int_{-\infty}^{\infty} {\rm d}\varepsilon f(\varepsilon ,\mu, T) \\
        &\times {\rm Tr} \Bigg[ 
        J_{\alpha}^{\mathcal{O}} \delta(\varepsilon - H) v_\beta \dfrac{{\rm d} G^{+}(\varepsilon)}{{\rm d}\varepsilon}
        - J_{\alpha}^{\mathcal{O}} \dfrac{{\rm d} G^{-}(\varepsilon)}{{\rm d}\varepsilon} v_\beta  \delta(\varepsilon - H)
        \Bigg]
        , \nonumber
    \end{align}
    where $f(\varepsilon ,\mu ,T)$ is the Fermi-Dirac distribution function, $G^{\pm}(\varepsilon)=(\varepsilon-H\pm\eta)^{-1}$ represent the retarded and advanced Green's functions, respectively, with $\eta$ acting as an imaginary energy broadening, and $\Omega$ is the system volume.

    Direct evaluation of this formula via full diagonalization scales as an order $\mathcal{N}^3$, where $\mathcal{N}$ is the dimension of the Hilbert space, severely limiting the maximum treatable system size. 
    To overcome this limitation and calculate macroscopic transport properties in the presence of real-space disorder, we expand the spectral operators using the Chebyshev Polynomial Method \cite{Weisse2006,Garcia2015,KITE2020,Vidarte2026}.
    The Hamiltonian and energy variables are first rescaled into the dimensionless interval $[-1, 1]$ to ensure convergence, and are represented by $\tilde{H}$ and $\tilde{\varepsilon}$.
    The delta functions and Green's functions are then expanded in a series of Chebyshev polynomials of the first kind, $T_m(x)$. 
    To filter out unphysical Gibbs oscillations resulting from the truncation of the series at a finite moment $M$, we incorporate the kernel $g_m$. 
    The resulting formula reads
    \begin{align}
        \sigma_{\alpha\beta}^{\mathcal{O}} (\mu ,T)\approx \dfrac{4e^2\hbar}{\pi\Omega} \dfrac{4}{\Delta E^2} \int_{-1}^{1} {\rm d} \tilde\varepsilon \; \dfrac{f(\tilde\varepsilon ,\tilde\mu, \tilde T)}{(1-\tilde\varepsilon)^2}    \zeta_{\alpha\beta}^{\mathcal{O}} (\tilde\varepsilon),
    \end{align}
    with 
    \begin{align}
        \zeta_{\alpha\beta}^{\mathcal{O}} (\tilde\varepsilon) \equiv \sum_{m,n}^M  
        \dfrac{g_mg_n}{(1+\delta_{m0})(1+\delta_{n0})}
        \Gamma_{m,n}(\tilde\varepsilon) \mu_{mn}^{\alpha\beta ,\mathcal{O}},
    \end{align}
    where
    \begin{align}
        \Gamma_{m,n}(\tilde\varepsilon)\equiv & 
        (\tilde\varepsilon - in \sqrt{1-\tilde\varepsilon^2}) e^{in\; {\rm arcos}(\tilde\varepsilon)} T_m(\tilde\varepsilon)  \nonumber \\ 
        &+(\tilde\varepsilon + im \sqrt{1-\tilde\varepsilon^2}) e^{-im\; {\rm arcos}(\tilde\varepsilon)} T_n(\tilde\varepsilon), 
    \end{align}
    \begin{align}
        \mu_{mn}^{\alpha\beta ,\mathcal{O}} \equiv {\rm Tr}\left[ T_m (\tilde H)J_{\alpha}^{\mathcal{O}}T_n (\tilde H) v_\beta\right],
    \end{align} 
    and $\Delta E\equiv E^+ - E^-$ is the spectral bandwidth, where $E^-$ and $E^+$ are the minimum and maximum eigenvalues.
    
    This formulation allows us to express the conductivity tensor as a sum over a two-dimensional matrix of Chebyshev moments $\mu_{mn}^{\alpha\beta ,\mathcal{O}}$.
    The computation of these moments relies exclusively on sequential sparse matrix-vector multiplications, which scale linearly with both the system size $\mathcal{N}$ and the number of expansion moments $M$.

    We implement this real-space approach using Quantum KITE \cite{KITE2020}, a high-performance open-source software designed for large-scale tight-binding simulations. 
    By using KITE’s optimized parallelization and memory-efficient algorithms, we calculate the continuous energy dependence of $\sigma_{\alpha\beta}^\mathcal{O}$ across the entire spectrum.  This setup enables us to efficiently account for millions of atomic sites, making it an ideal tool for investigating the impacts of realistic, random structural defects without resorting to artificial supercells.

\section{Results and Discussion}
\label{Sec:results}
We consider three complementary disorder mechanisms acting on the full \(sp^3\) lattice model: vacancies, orbital-dependent Anderson  on-site disorder, and \(V_{sp\sigma}\)-bond disorder. 
They are introduced in this order to separate different ways of perturbing the orbital texture. 
Vacancies remove complete lattice sites and all orbital channels attached to them; orbital-dependent Anderson disorder shifts the on-site energies of different orbitals independently, producing random local crystal-field environments; and \(V_{sp\sigma}\)-bond disorder acts directly on the \(s\)-\(p\) hybridization channel. 
A fourth case, dilute controlled on-site impurities, is treated separately in Appendix~\ref{App:on-site_disorder}. 
Each disordered calculation is compared with the pristine limit. 
The same Quantum KITE implementation is used to compute both bulk transport coefficients and momentum-/real-space orbital-resolved spectral quantities, with the pristine spectral function and quadrupole textures validated against exact diagonalization in Appendix~\ref{App:kite-arpes}. 
Below we focus on the transport coefficients -- $\sigma^{L_z}_{\rm OHE}$, $\sigma_{xx}$, and $\theta_{\rm OH}$ -- for the three disorder mechanisms. The corresponding momentum-resolved spectral functions and quadrupole textures are collected in Appendix~\ref{App:disorder-textures}, where they provide a microscopic counterpart to the transport trends while keeping the main discussion focused on transport.
All calculations reported below use identical simulation parameters across every disorder type studied~\cite{simparams}.

\subsection{Vacancy disorder}
\label{SubSec:vacancy}

    \begin{figure}[t]
        \centering
        \includegraphics[width=1.0\linewidth]{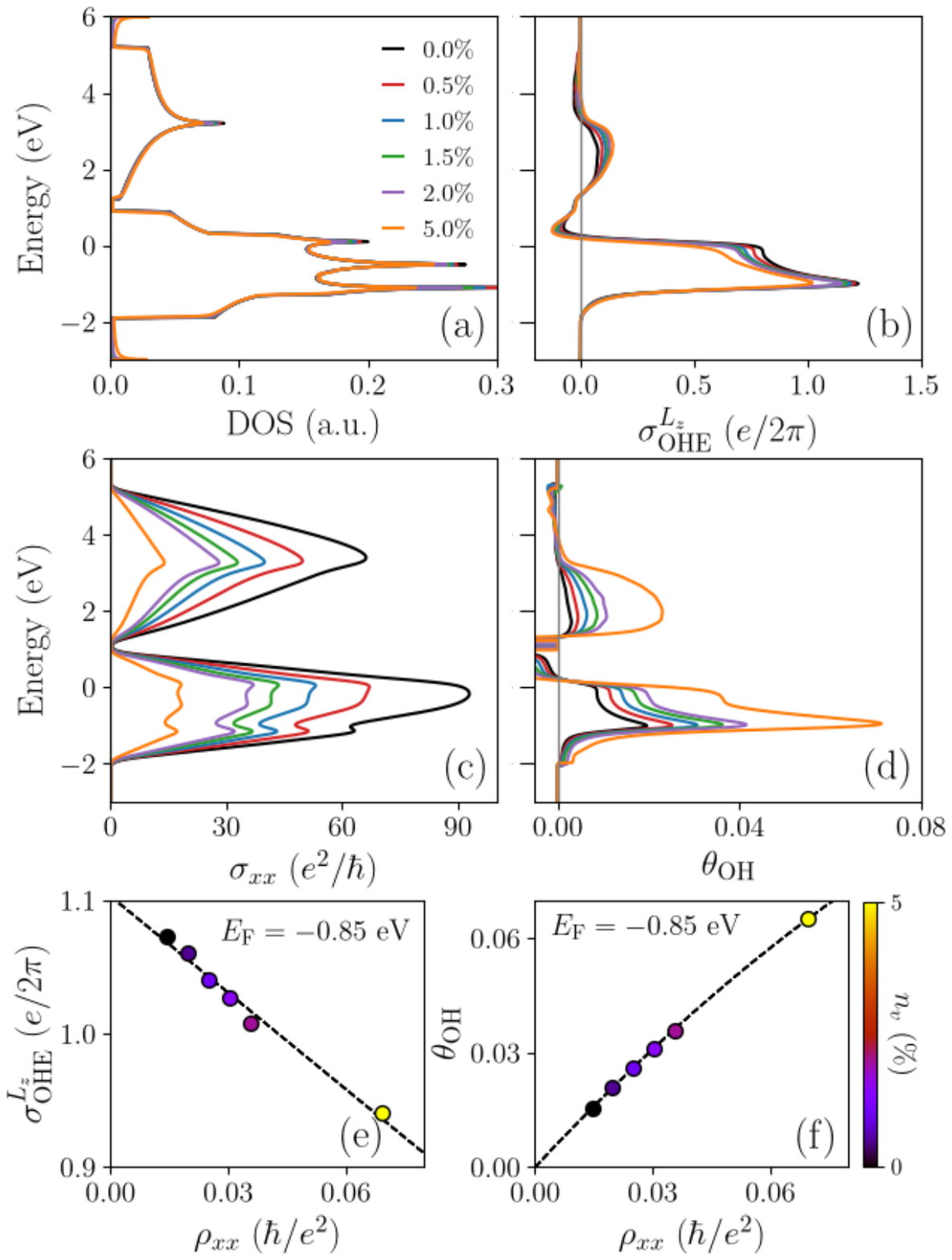}
        \caption{
        Vacancy disorder. 
        (a) Density of states, (b) orbital Hall conductivity $\sigma^{L_z}_{\rm OHE}$, (c) longitudinal conductivity $\sigma_{xx}$, and (d) orbital Hall angle $\theta_{\rm OH}$ as a function of energy, for vacancy concentrations $n_v\lesssim5.0\%$. 
        The black line corresponds to the pristine system.
        (e) $\sigma^{L_z}_{\rm OHE}$ and (f) $\theta_{\rm OH}$ at fixed $E_F=-0.85$~eV plotted against $\rho_{xx}$, parameterized by $n_v$.
        }
        \label{fig:Fig_04}
    \end{figure}

    We begin with vacancy disorder, which provides a simple model of structural disorder.
    A subset of lattice sites is randomly selected with concentration \(n_v\), and these sites are removed from the lattice by setting all hopping amplitudes connected to them to zero,
    \(H \rightarrow H_{\rm vac}=P_V H P_V\), where \(P_V\) projects onto the subspace of non-vacant sites.

    Figure~\ref{fig:Fig_04} summarizes the resulting transport behavior for vacancy concentrations up to \(n_v=5\%\).
    Panels~(a)--(d) show the disorder-averaged density of states, orbital Hall conductivity \(\sigma^{L_z}_{\rm OHE}\), longitudinal conductivity \(\sigma_{xx}\), and orbital Hall angle
    \(\theta_{\rm OH}=\sigma^{L_z}_{\rm OHE}/\sigma_{xx}\) as a function of energy.
    Increasing the vacancy concentration suppresses both \(\sigma_{xx}\) and \(\sigma^{L_z}_{\rm OHE}\), but the two responses are affected at different rates.
    Over the energy range shown, the longitudinal conductivity is strongly reduced, while the orbital Hall conductivity remains comparatively robust.

    To make the comparison more quantitative, we focus on \(E_F=-0.85\,{\rm eV}\), an energy close to the orbital Hall peak of the pristine model and in the same energy window commonly used in previous studies of the \(sp^3\) OHE model~\cite{Go2018}.
    Figures~\ref{fig:Fig_04}(e) and \ref{fig:Fig_04}(f) show \(\sigma^{L_z}_{\rm OHE}\) and \(\theta_{\rm OH}=\sigma^{L_z}_{\rm OHE}/\sigma_{xx}\) at this fixed Fermi energy as a function of the longitudinal resistivity, \(\rho_{xx}=1/\sigma_{xx}\), with the color scale indicating the vacancy concentration.
    Using \(\rho_{xx}\) rather than \(n_v\) on the horizontal axis allows us to compare different disorder mechanisms through their effect on charge transport, and also makes contact with experiments where disorder is often discussed in terms of resistivity. As shown in Figs.~\ref{fig:Fig_04}(e) and \ref{fig:Fig_04}(f), \(\sigma^{L_z}_{\rm OHE}\) decreases approximately linearly with \(\rho_{xx}\) over the range studied, while \(\theta_{\rm OH}\) increases. 
    In this range, the increase of \(\rho_{xx}\) dominates over the moderate decrease of \(\sigma^{L_z}_{\rm OHE}\), leading to a nearly linear increase of the orbital Hall angle. This trend is similar to the increase of orbital Hall angle with resistivity reported in Ru\(_{1-x}\)Cu\(_x\) and Ru\(_{1-x}\)Ti\(_x\) alloy bilayers~\cite{Yang2026}.
    The corresponding momentum-resolved spectral function and quadrupole-texture response are shown in Appendix~\ref{App:disorder-textures}, Fig.~\ref{fig:Fig_05}, where the main quadrupolar character remains visible as the spectral function broadens.

\subsection{Anderson Disorder}
\label{SubSec:Anderson}

    We next consider Anderson disorder implemented with orbital-dependent on-site shifts,
    \[
    H \to H + \sum_{i,\alpha} w_i^\alpha\, c^\dagger_{i\alpha}c_{i\alpha},
    \] 
    where \(w_i^\alpha\) is drawn independently for each site \(i\) and orbital \(\alpha\) from a uniform distribution
    \(w_i^\alpha \in [-W/2,W/2]\).
    This differs from the usual scalar Anderson disorder, in which all orbitals on the same site would be shifted by the same random potential. Here, the disorder also changes the relative on-site energies of the \(s\), \(p_x\), \(p_y\), and \(p_z\) orbitals. As a result, it introduces random local crystal-field variations and modifies the local orbital-hybridization environment, in addition to ordinary site-diagonal disorder. Thus, although the hopping amplitudes themselves are unchanged, the disorder changes the local energy offsets that control \(s\)-\(p\) mixing and the quadrupolar texture of the clean model.

    \begin{figure}[h]
        \centering
        \includegraphics[width=1.0\linewidth]{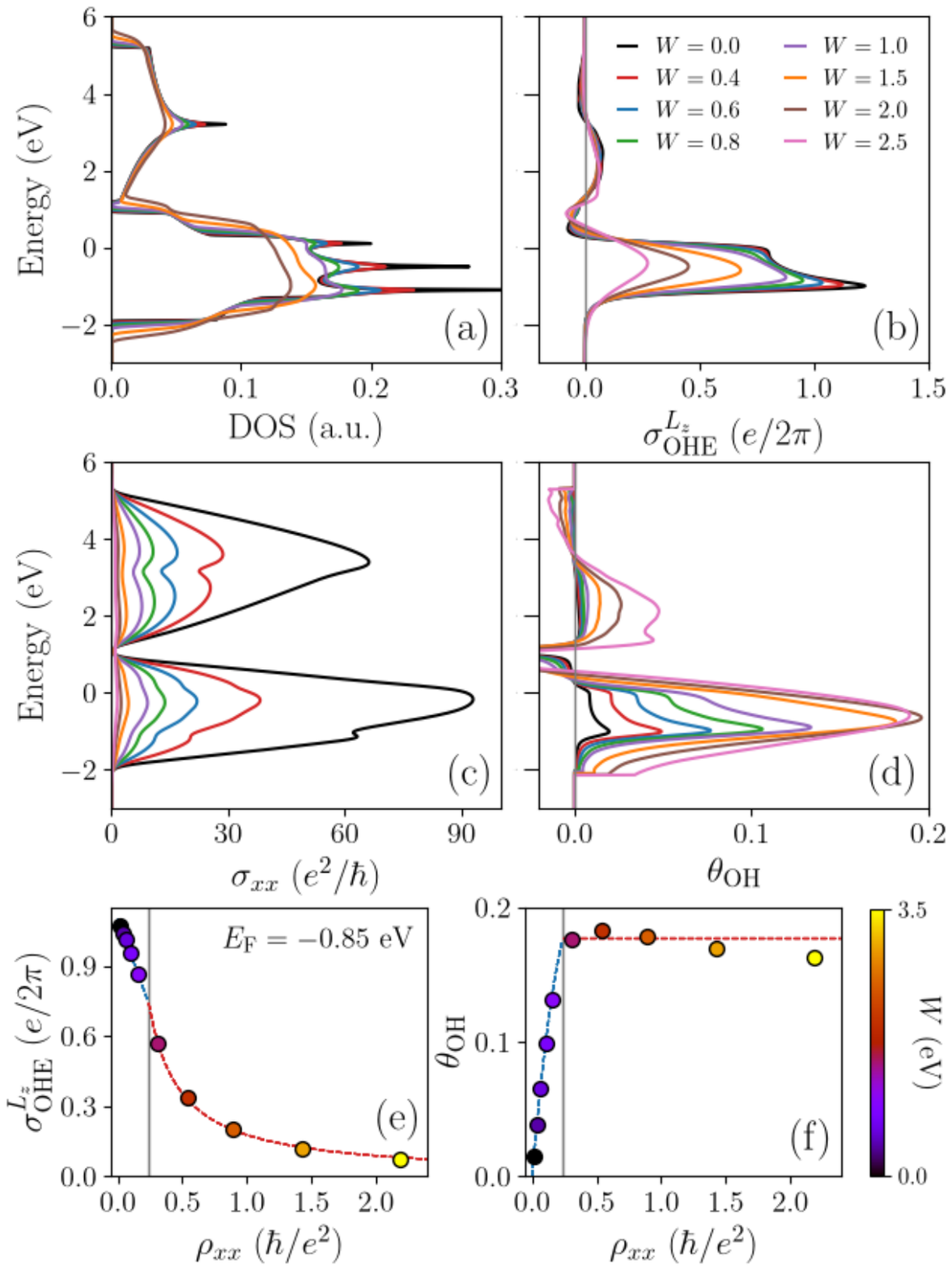}
        \caption{
        Anderson disorder. 
        (a) Density of states, 
        (b) orbital Hall conductivity $\sigma^{L_z}_{\rm OHE}$, 
        (c) longitudinal conductivity $\sigma_{xx}$, and 
        (d) orbital Hall angle $\theta_{\rm OH}=\sigma^{L_z}_{\rm OHE}/\sigma_{xx}$ as a function of energy, for disorder strengths $W\lesssim 2.5$~eV.
        The black line corresponds to the pristine system.
        {(e) $\sigma^{L_z}_{\rm OHE}$ and (f) $\theta_{\rm OH}$ at fixed $E_F=-0.85$~eV plotted against the longitudinal resistivity $\rho_{xx}=1/\sigma_{xx}$, parameterized by $W$ up to $3.5$~eV.}
        }
        \label{fig:Fig_02}
    \end{figure}

    Figure~\ref{fig:Fig_02} summarizes the transport behavior for Anderson disorder implemented with orbital-dependent on-site shifts.
    Panels~(a)--(d) show the disorder-averaged density of states,  \(\sigma^{L_z}_{\rm OHE}\),\(\sigma_{xx}\), and orbital Hall angle \(\theta_{\rm OH}\)  as a function of energy, for disorder strengths up to \(W=2.5\)~eV.
    Panels~(e) and (f) extend the analysis to \(W=3.5\)~eV and show \(\sigma^{L_z}_{\rm OHE}\) and \(\theta_{\rm OH}\) at fixed Fermi energy \(E_F=-0.85\)~eV as a function of the longitudinal resistivity.
    
    Increasing \(W\) broadens the spectral features and suppresses both \(\sigma_{xx}\) and \(\sigma^{L_z}_{\rm OHE}\), but the effect on the longitudinal conductivity is much stronger.
    Compared with vacancies, Anderson disorder gives access to a broader range of \(\rho_{xx}\), allowing us to follow the evolution from the weak-to-moderate disorder regime to a more strongly disordered regime.
    At low and intermediate resistivities, \(\sigma^{L_z}_{\rm OHE}\) decreases approximately linearly with \(\rho_{xx}\), similar to the vacancy case, $\sigma^{L_z}_{\rm OHE} \approx \sigma_0-\gamma_w\rho_{xx}$,
    with \(\sigma_0=1.102~e/2\pi\) and \(\gamma_w=1.512~\hbar/2\pi e\). At larger resistivities, the behavior changes: \(\sigma^{L_z}_{\rm OHE}\) crosses over to an approximately inverse dependence on \(\rho_{xx}\), $$
    \sigma^{L_z}_{\rm OHE} \approx \frac{C}{\rho_{xx}}$$ with \(C=0.1775~e/2\pi\).
    Since \(\theta_{\rm OH}=\sigma^{L_z}_{\rm OHE}\rho_{xx}\), this second regime corresponds to the saturation of the orbital Hall angle, as seen in Fig.~\ref{fig:Fig_02}(f). The crossover between the two regimes occurs around \(\rho_{xx}\approx0.24~\hbar/e^2\).
    
    The resulting behavior connects with recent experiments on orbital Hall systems with tunable disorder.
    At weak to moderate disorder, the increase of \(\theta_{\rm OH}\) with resistivity is similar to the trend reported in Ru\(_{1-x}\)Cu\(_x\) and Ru\(_{1-x}\)Ti\(_x\) alloy bilayers~\cite{Yang2026}.
    At stronger disorder, \(\theta_{\rm OH}\) approaches an approximately constant value, which means that \(\sigma^{L_z}_{\rm OHE}\) scales with \(\sigma_{xx}\). This is the same type of scaling reported in Mn thin films spanning amorphous to single-crystalline order~\cite{Kang2026}.
    The corresponding momentum-resolved spectral function and quadrupole-texture response are shown in Appendix~\ref{App:disorder-textures}, Fig.~\ref{fig:Fig_03}; as for vacancies, the dominant quadrupolar character remains identifiable as the spectral function broadens.

\subsection{$V_{sp\sigma}$ Bond Disorder}
\label{SubSec:bond-disorder}

    Finally, we consider a disorder mechanism that acts directly on the \(s\)-\(p\) hybridization responsible for the orbital Hall response in the clean model.
    A random subset of sites, with fraction \(n_b\), is selected, and at every selected site the four \(V_{sp\sigma}\) bonds touching it -- along \(+\hat{x}\), \(-\hat{x}\), \(+\hat{y}\), and \(-\hat{y}\) -- are set to zero.
    All other hopping amplitudes, \(V_{ss\sigma}\), \(V_{pp\sigma}\), and \(V_{pp\pi}\), are kept unchanged.
    This construction locally suppresses the \(s\)-\(p\) mixing, and therefore the hybridization channel that generates the quadrupolar texture and the orbital Berry curvature, without removing the remaining hopping network.
    Charge transport can still proceed through the \(s\)-\(s\) and \(p\)-\(p\) channels, while the orbital Hall response is perturbed more selectively through the \(s\)-\(p\) sector.
    
    \begin{figure}[t]
        \centering
        \includegraphics[width=1.0\linewidth]{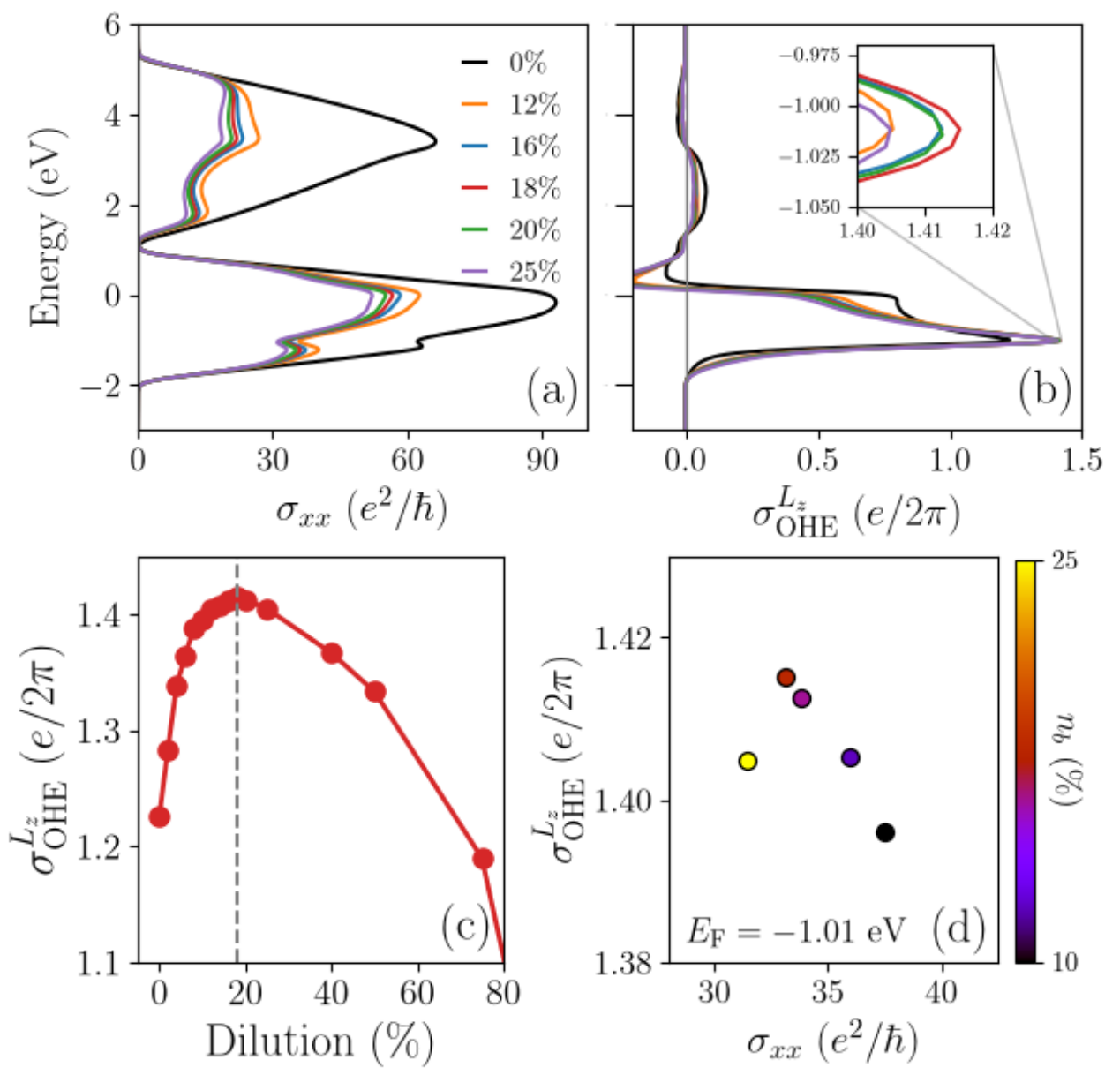}
        \caption{
        $V_{sp\sigma}$ bond disorder. 
        (a) Longitudinal conductivity $\sigma_{xx}$ and (b) orbital Hall conductivity $\sigma^{L_z}_{\rm OHE}$ as a function of energy, for dilution fraction $n_b=0$-$25\%$; inset in (b) zooms into the curve crossings near $\varepsilon\approx-1$~eV. 
        (c) $\sigma^{L_z}_{\rm OHE}$ at fixed $E_F=-1.01$~eV against $n_b$, showing a non-monotonic peak near $n_b\approx18\%$. 
        (d) Same data as $\sigma^{L_z}_{\rm OHE}$ against $\sigma_{xx}$, parameterized by $n_b$.
        }
        \label{fig:Fig_06}
    \end{figure}

    Equation~\eqref{eq:omega_final} shows that, in the clean low-energy model, \(\Omega^{L_z}_{+,yx}\propto\mathcal{A}C\mathcal{B}\), with \(C\propto V_{sp\sigma}^2\).
    Thus, continuously tuning \(V_{sp\sigma}\to0\) suppresses the orbital Hall response.
    This limiting behavior was previously demonstrated on a cubic lattice using the same \(sp^3\) framework~\cite{Go2018}; the clean full-lattice dependence on \(V_{sp\sigma}\) for the present square lattice is analyzed in Appendix~\ref{App:OHE-pristine}.
    Here, instead of globally rescaling the hopping parameter, we ask what happens when the \(s\)-\(p\) hybridization is removed locally and stochastically, while the rest of the hopping network remains intact.

    Figure~\ref{fig:Fig_06} shows the transport response for \(V_{sp\sigma}\)-bond dilution with \(n_b=0\)--\(25\%\), where \(n_b=0\) corresponds to the pristine system.
    Panels~(a) and (b) show \(\sigma_{xx}\) and \(\sigma^{L_z}_{\rm OHE}\) as functions of energy.
    Since only the \(s\)-\(p\) hopping is diluted, while the \(s\)-\(s\) and \(p\)-\(p\) hoppings remain intact, the longitudinal conductivity is only moderately affected over this range.
    The orbital Hall conductivity, by contrast, is modified more directly because the \(s\)-\(p\) hybridization is the channel that generates the quadrupolar texture and the orbital Berry curvature in the clean model.

    The inset in Fig.~\ref{fig:Fig_06}(b) highlights the crossings of the \(\sigma^{L_z}_{\rm OHE}(\varepsilon)\) curves near \(\varepsilon\simeq -1\)~eV, which motivates the choice of \(E_F=-1.01\)~eV for the fixed-energy analysis.
    At this Fermi energy, Fig.~\ref{fig:Fig_06}(c) shows a non-monotonic dependence on the dilution fraction: \(\sigma^{L_z}_{\rm OHE}\) first increases with \(n_b\), reaches a maximum near \(n_b\simeq18\%\), and then decreases as the \(s\)-\(p\) hybridization is further diluted.
    Panel~(d) shows the same data plotted against \(\sigma_{xx}\), with \(n_b\) as the tuning parameter. 

    To understand this behavior, we can resort to the low-energy model and ask how the clean orbital Hall response depends on \(V_{sp\sigma}\). 
    In this model, Eq.~\eqref{eq:omega_final} shows that the Berry curvature is controlled by the product \(\mathcal{A}C\mathcal{B}\), with \(C=4a^2V_{sp\sigma}^2/(\varepsilon_p-\varepsilon_s)\), \(\mathcal{A}=A+C/2\), and \(\mathcal{B}=B+C/2\). 
    The dependence on \(V_{sp\sigma}\) is therefore not monotonic in general. 
    Increasing \(V_{sp\sigma}\) increases the magnitude of \(C\), which is the coefficient associated with the \(Q_{xy}\)-like texture generated by virtual \(p\)-\(s\)-\(p\) processes. 
    At the same time, for the parameters used here, \(C<0\), so the same hybridization reduces both \(\mathcal{A}\) and \(\mathcal{B}\), including the coefficient associated with the \(Q_{x^2-y^2}\)-like texture. 
    The orbital Hall response is then maximized not by making the \(s\)-\(p\) hybridization as large as possible, but by balancing the two quadrupolar channels: the \(Q_{xy}\) texture must be sizable, while the \(Q_{x^2-y^2}\) texture and the orbital-current prefactor remain sufficiently large.

    This interpretation is supported by the clean full-lattice calculation reported in Appendix~\ref{App:OHE-pristine}. 
    There, \(\sigma^{L_z}_{\rm OHE}\) is computed for several uniformly rescaled values of \(V_{sp\sigma}\), and the response is already non-monotonic in the pristine system near \(E_F\simeq -1\) eV. 
    The precise position of the maximum depends on energy: for example, the maximum occurs near \(V_{sp\sigma}\simeq0.50\) eV at \(E_F=-1.01\) eV, while it shifts toward \(V_{sp\sigma}\simeq0.55\) eV at the nearby energy \(E_F=-1.035\) eV. 
    The low-energy estimate should therefore be viewed as a guide to the underlying hybridization balance, rather than as a quantitative prediction for the full lattice. 
    If the stochastic bond dilution is interpreted only as a crude effective reduction of the hybridization, \(V_{sp\sigma}^{\rm eff}\simeq V_{sp\sigma}^{0}(1-n_b)^2\), the observed maximum near \(n_b\simeq18\%\) gives \(V_{sp\sigma}^{\rm eff}\simeq0.336\) eV, close to the small-\(k\) stationary point \(V_{sp\sigma}\simeq0.348\) eV. 
    This agreement should not be read as an exact mapping between random bond dilution and a clean hopping renormalization. 
    Bond dilution also modifies the local electronic environment around the removed bonds, redistributes spectral weight, broadens spectral features, and mixes momenta, which can shift the maximum relative to the clean estimate. 
    Rather, the comparison indicates that the non-monotonic clean dependence on the hybridization is one ingredient of the bond-dilution peak, while the numerical calculation retains the disorder-specific effects absent from the estimate.
    The corresponding momentum-resolved spectral function and quadrupole-texture response are shown in Appendix~\ref{App:disorder-textures}, Fig.~\ref{fig:Fig_07}; bond dilution redistributes the $Q_{xy}$ texture while the $Q_{x^2-y^2}$ sign structure remains recognizable at the transport maximum.

\subsection{Additional disorder and real-space results}

The main text has focused on vacancies, orbital-dependent Anderson disorder, and \(V_{sp\sigma}\)-bond dilution. 
Two additional sets of calculations, reported in the Appendices, complement this analysis.

Appendix~\ref{App:on-site_disorder} considers dilute controlled on-site impurities, where a fixed fraction of sites receives the same energy shift \(\delta U\). 
This case is useful because it separates dilute impurity physics from the fully random, orbital-dependent Anderson disorder discussed in Sec.~\ref{SubSec:Anderson}. 
The main result is that positive and negative impurity potentials of the same magnitude do not affect the transport coefficients in the same way. 
This asymmetry is consistent with resonant impurity scattering: the effect depends on where the impurity-induced spectral feature lies relative to the Fermi energy, not only on the magnitude of \(\delta U\).
Although $\delta U$ shifts all orbitals on an impurity site by the same amount, it changes their energy detuning relative to orbitals on neighboring host sites and can therefore modify the local $s$-$p$ hybridization in addition to producing resonant scattering.

Appendix~\ref{App:real-space} complements the momentum-space spectral analysis with real-space maps of the LDOS and local orbital multipoles. 
These maps clarify an important distinction between orbital quadrupoles and orbital angular momentum in the present spinless model. 
The quadrupolar operators \(Q_{xy}\) and \(Q_{x^2-y^2}\) are real orbital operators, so local disorder can generate finite real-space quadrupolar patterns by breaking the spatial cancellation present in the pristine lattice. 
By contrast, \(L_z\) is purely imaginary in the real \(p_x,p_y\) orbital basis, and all disorder potentials considered here preserve the reality of the Hamiltonian. 
The eigenstates can therefore still be chosen real, forcing the local expectation value \(\langle L_z\rangle\) to vanish site by site, not only after spatial averaging. 
Thus, disorder creates local orbital anisotropy, visible in the quadrupolar textures, but it does not create a static local orbital chirality in this model.
\section{Conclusions}
\label{Sec:conclusions}

We have studied how different forms of disorder affect the orbital Hall response of a centrosymmetric spinless \(sp^3\) tight-binding model on a square lattice. 
The clean model provides the microscopic reference for the analysis: \(s\)-\(p\) hybridization generates quadrupolar orbital textures in the \(p_x,p_y\) sector, and the low-energy theory relates the projected orbital Berry curvature to the product \(\mathcal{A}C\mathcal{B}\). 
This shows that the orbital Hall response is controlled by the interplay between the \(Q_{xy}\) and \(Q_{x^2-y^2}\) channels, rather than by a single orbital texture or by the magnitude of \(V_{sp\sigma}\) alone.

For the more generic forms of disorder studied here, vacancies and orbital-dependent Anderson disorder, the results are most clearly organized in terms of the longitudinal resistivity \(\rho_{xx}\). 
Both disorder mechanisms increase \(\rho_{xx}\) and suppress the longitudinal conductivity more strongly than the orbital Hall conductivity over the range studied. 
As a result, the orbital Hall angle increases with \(\rho_{xx}\). 
In the Anderson case, where a broader resistivity range is accessible, this increase crosses over to a saturation regime, corresponding to an approximately linear scaling between \(\sigma^{L_z}_{\rm OHE}\) and \(\sigma_{xx}\). 
Thus, although vacancies and Anderson disorder act differently at the microscopic level, they support the same qualitative conclusion: the orbital Hall response remains sizable even when charge transport is strongly degraded.

The momentum-resolved spectral functions provide a microscopic counterpart to this transport trend. 
Vacancy and Anderson disorder broaden the bands and produce weak texture leakage away from the clean-limit symmetry directions, but the dominant quadrupolar character of the main bands remains visible. 
The persistence of the dominant quadrupolar texture is consistent with the transport observation that \(\sigma^{L_z}_{\rm OHE}\) is suppressed more slowly than \(\sigma_{xx}\), allowing \(\theta_{\rm OH}\) to increase as the system becomes more resistive.

Bond dilution behaves differently because it does not simply act as a generic source of resistivity. 
It selectively modifies the \(s\)-\(p\) hybridization channel that generates the \(Q_{xy}\)-like texture and contributes to the orbital Berry curvature. 
Instead of a monotonic suppression, the orbital Hall conductivity develops a maximum near \(n_b\simeq18\%\). 
This behavior is consistent with the clean low-energy analysis: reducing the effective \(s\)-\(p\) hybridization can increase the response over an intermediate range because the Berry curvature depends on the balance between \(\mathcal{A}\), \(C\), and \(\mathcal{B}\). 
In the disordered calculation this tuning is achieved by stochastic bond removal rather than by a uniform hopping renormalization, so the peak combines the clean hybridization balance with disorder-induced broadening and momentum mixing.

The additional calculations reported in the Appendices extend this picture beyond the three main disorder mechanisms. 
Dilute controlled on-site impurities reveal a sign-dependent response: attractive and repulsive impurity potentials of the same magnitude affect the transport coefficients differently, consistent with resonant scattering and with the fact that the impurity potential changes the local intersite $s$-$p$ detuning relative to neighboring host sites.
Real-space texture maps show that disorder generates local quadrupolar patterns around the perturbed regions, while the local orbital angular momentum remains zero site by site. 
This distinction reflects the reality of the spinless Hamiltonian: the quadrupolar operators are real and can acquire local expectation values when spatial cancellation is broken, whereas \(L_z\) is purely imaginary in the real \(p_x,p_y\) basis.

Overall, our results show that disorder affects orbital Hall transport in two complementary ways. 
When disorder mainly increases \(\rho_{xx}\), as in the vacancy and Anderson cases, the orbital Hall conductivity is comparatively robust and the orbital Hall angle grows with resistivity, eventually reaching a saturation regime in the strongly disordered Anderson case. 
When disorder selectively modifies the \(s\)-\(p\) hybridization, as in bond dilution, it can tune the quadrupolar texture itself and enhance the orbital Hall response over an intermediate range. 
The orbital Hall response of the \(sp^3\) model is therefore not controlled by disorder strength alone, but by the combined effect of resistivity, spectral broadening, and the microscopic orbital channel that is being perturbed.

\begin{acknowledgments}
TGR, acknowledges the  FCT - Fundação para a Ciência e Tecnologia for financial support (Grant No. 2023.11755.PEX and DOI identifier https://doi.org/10.54499/2023.11755.PEX)
TGR acknowledges  FCT  project reference numbers UIDB/04650/2020,  and 2022.07471.CEECIND/CP1718/CT0001
(with DOI identifier: 10.54499/2022.07471.CEECIND/CP1718/
CT0001) and CNPq (Grant No 305013/2024-6). 
TGR and KJUV acknowledge support from the EIC Pathfinder OPEN grant 101129641 “OBELIX”. TPC acknowledges CNPq (Grant No. 305647/2024-5). 
HPV acknowledges the University of York High-Performance Computing service, Viking.
Further support from FCT through the PhD Grant. No. 2024.00560.BD (HPV) is acknowledged.

\end{acknowledgments}

\appendix

\section{Orbital-resolved spectral functions using Quantum KITE} 
\label{App:kite-arpes}
\setcounter{figure}{0}
\renewcommand{\thefigure}{A\arabic{figure}}

Beyond the linear-response conductivities discussed in Sec.~\ref{Sec:theory&model}, Quantum KITE~\cite{KITE2020} also gives access to momentum- and orbital-resolved spectral quantities. 
Two target functions are used throughout this Appendix, \texttt{arpes()} and \texttt{spectral\_map()}.
The first evaluates the spectral function $A(\mathbf{k},\varepsilon)$ along a user-specified path in the Brillouin zone.
The second evaluates the spectral function over the \emph{full} two-dimensional Brillouin zone at a single fixed energy.
{Both target functions accept a \texttt{weight} argument that assigns a coefficient to each orbital in the unit cell, allowing the spectral trace to be projected onto an arbitrary coherent combination of orbitals rather than onto a single orbital.}
For example, using the basis $\{|s\rangle,|p_x\rangle,|p_y\rangle,|p_z\rangle\}$ from our Hamiltonian, Eq.~\eqref{eq:real-sapce_hamiltonian}, the \texttt{weight} parameter lets target functions isolate the eigenstates of each orbital multipole individually: 
$\texttt{weight}=(0,1,\pm1,0)$ projects onto the $Q_{xy}=\pm1$ eigenstates $(|p_x\rangle\pm|p_y\rangle)/\sqrt2$; $\texttt{weight}=(0,1,0,0)$ or $(0,0,1,0)$ projects onto the $Q_{x^2-y^2}=\pm1$ eigenstates, which are simply the bare $|p_x\rangle,|p_y\rangle$ orbitals; 
and $\texttt{weight}=(0,1,\pm i,0)$ projects onto the $L_z=\pm\hbar$ orbital-dipole eigenstates $(|p_x\rangle\pm i|p_y\rangle)/\sqrt2$. 
This is what allows KITE to resolve the orbital-quadrupole and orbital-dipole textures separately in momentum space, in addition to the ordinary spectral function.

\begin{figure}[t]
    \centering
    \includegraphics[width=1.00\linewidth]{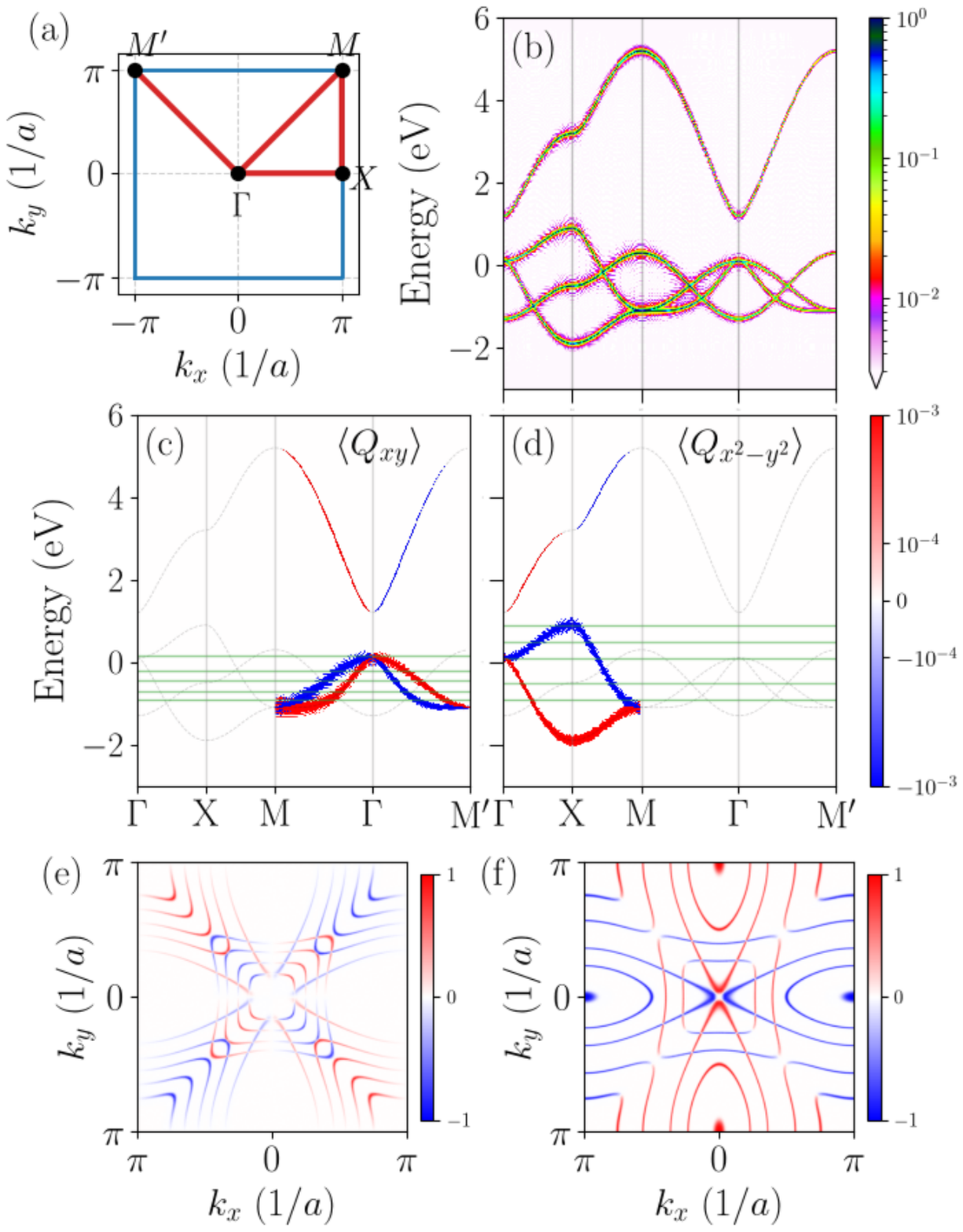}
    \caption{
    Spectral properties and texture maps calculated via Quantum KITE from the full lattice Hamiltonian. 
    (a)~Square-lattice Brillouin zone showing the $\Gamma$--$\mathrm{X}$--$\mathrm{M}$--$\Gamma$--$\mathrm{M}'$ path. 
    (b)~Spectral function $A(\mathbf{k},\varepsilon)$ along the path. 
    (c),~(d)~Expectation values $\langle Q_{xy}\rangle$ and $\langle Q_{x^2-y^2}\rangle$ resolved along the path. 
    (e),~(f)~Corresponding textures projected onto the full Brillouin zone at fixed energies indicated by the green lines in panels~(c) and~(d), respectively.
    }
    \label{fig:appendixA}
\end{figure}

Figure~\ref{fig:appendixA}(b) shows the resulting spectral function $A(\mathbf{k},\varepsilon)$; 
panels (c) and (d) show $\langle Q_{xy}\rangle$ and $\langle Q_{x^2-y^2}\rangle$ resolved along the same path, and panels (e) and (f) show both textures projected onto the full Brillouin zone at fixed energy $\varepsilon=0.15$~eV (dashed line in (b)-(d)). 
Note the path revisits $\Gamma$ before continuing to $M'$, the same Brillouin-zone corner as $M$ reached via the opposite diagonal ($k_x=-k_y$ instead of $k_x=k_y$): 
since $\langle Q_{xy}\rangle\propto k_xk_y$ changes sign between the two diagonals, this choice of path makes the fourfold quadrupole texture of Fig.~\ref{fig:appendixA}(c) directly visible, rather than hidden by the point-group equivalence of $M$ and $M'$.
To validate this approach, we compare KITE's output against the exact diagonalization of the full lattice Hamiltonian of Sec.~\ref{SubSec:model_Hamiltonian} — i.e., without the quadratic Löwdin truncation used to obtain Eq.~\eqref{eq:effective_hamiltonian} — along the path shown in Fig.~\ref{fig:appendixA}(a).

\section{Momentum-Resolved Orbital-Texture Response to Disorder}
\label{App:disorder-textures}
\setcounter{figure}{0}
\renewcommand{\thefigure}{B\arabic{figure}}

    This appendix collects the momentum-resolved spectral functions and quadrupole-texture response for the three disorder mechanisms discussed in Sec.~\ref{Sec:results}, complementing the transport results shown there.

    \subsection*{Vacancy disorder}

    \begin{figure}[t]
    \centering
    \includegraphics[width=1.0\linewidth]{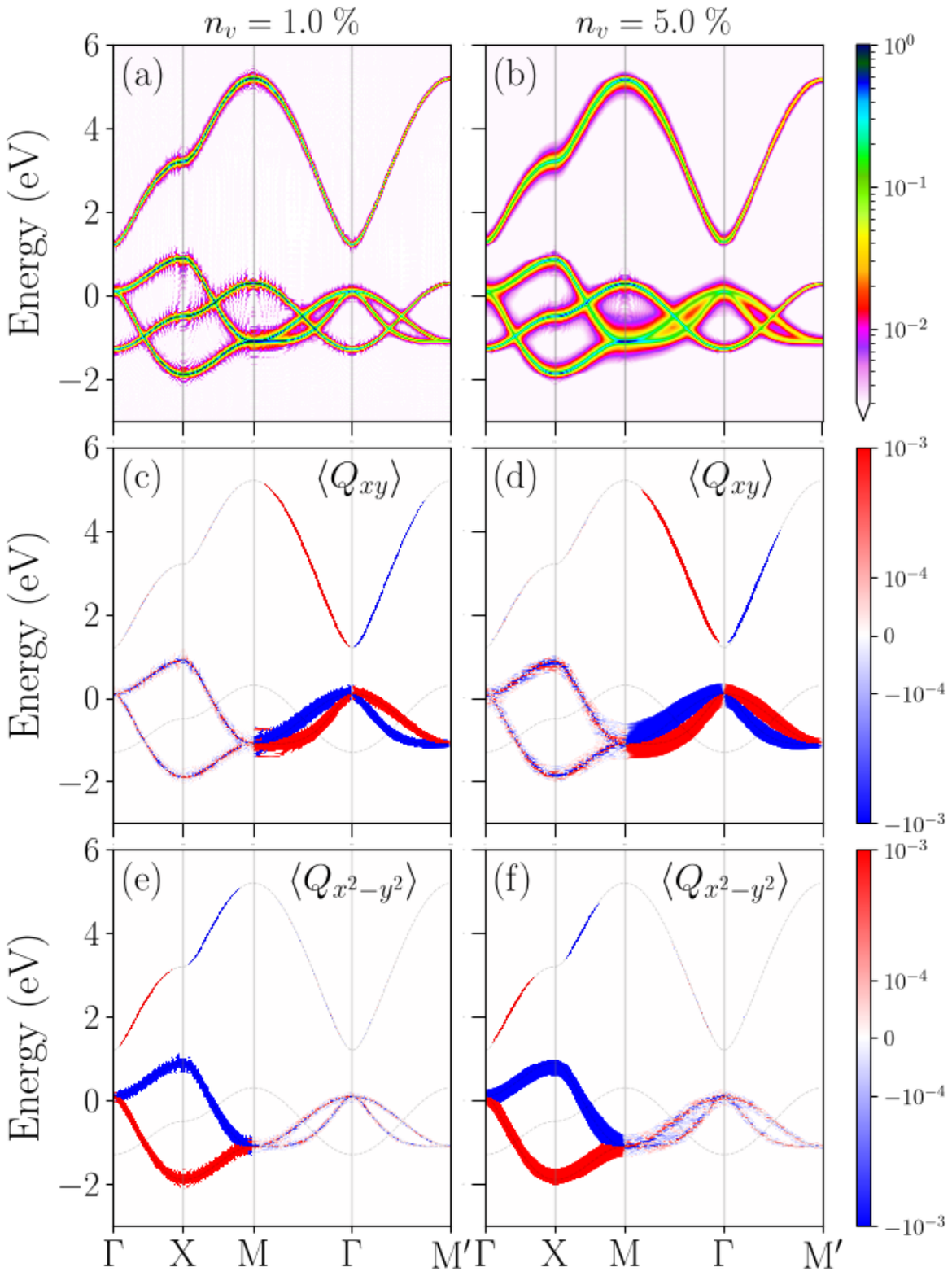}
    \caption{
    Vacancy disorder. 
    Spectral and orbital-texture properties for $n_v=1.0\%$ (left column) and $5.0\%$ (right column), along the $\Gamma$-${\rm X}$-${\rm M}$-$\Gamma$-${\rm M}'$ path of Fig.~\ref{fig:appendixA}(a). 
    (a),(b) Spectral function. 
    (c),~(d)~$\langle Q_{xy}\rangle$ and (e),~(f)~$\langle Q_{x^2-y^2}\rangle$, both plotted on a $[-10^{-3}, 10^{-3}]$ color scale.
    }
    \label{fig:Fig_05}
    \end{figure}

    Figure~\ref{fig:Fig_05} shows the corresponding spectral function and quadrupolar textures along the \(\Gamma-X-M-\Gamma-M'\) path, for \(n_v=1.0\%\) and \(5.0\%\).
    The spectral function broadens visibly as the vacancy concentration increases, reflecting the loss of translational invariance and the stronger momentum mixing.
    The quadrupolar textures, however, remain largely concentrated on the same main band branches as in the clean system.
    This is most apparent in the saturated red and blue features of \(\langle Q_{xy}\rangle\) and \(\langle Q_{x^2-y^2}\rangle\), which persist even at \(n_v=5.0\%\).
    
    Vacancy disorder also produces weak texture leakage into regions where the corresponding clean-limit quadrupole component vanishes: \(\langle Q_{xy}\rangle\) appears along \(\Gamma-X\), and \(\langle Q_{x^2-y^2}\rangle\) appears along \(M-\Gamma-M'\).
    This leakage increases with vacancy concentration, but remains small on the \([-10^{-3},10^{-3}]\) scale used in the figure.
    Thus, the spectral functions show clear disorder broadening, while the underlying quadrupolar character of the main bands is comparatively robust.
    This provides a microscopic counterpart to the transport result that \(\sigma^{L_z}_{\rm OHE}\) is less affected by vacancies than \(\sigma_{xx}\).

    \subsection*{Anderson disorder}


    \begin{figure}[t]
        \centering
        \includegraphics[width=1.0\linewidth]{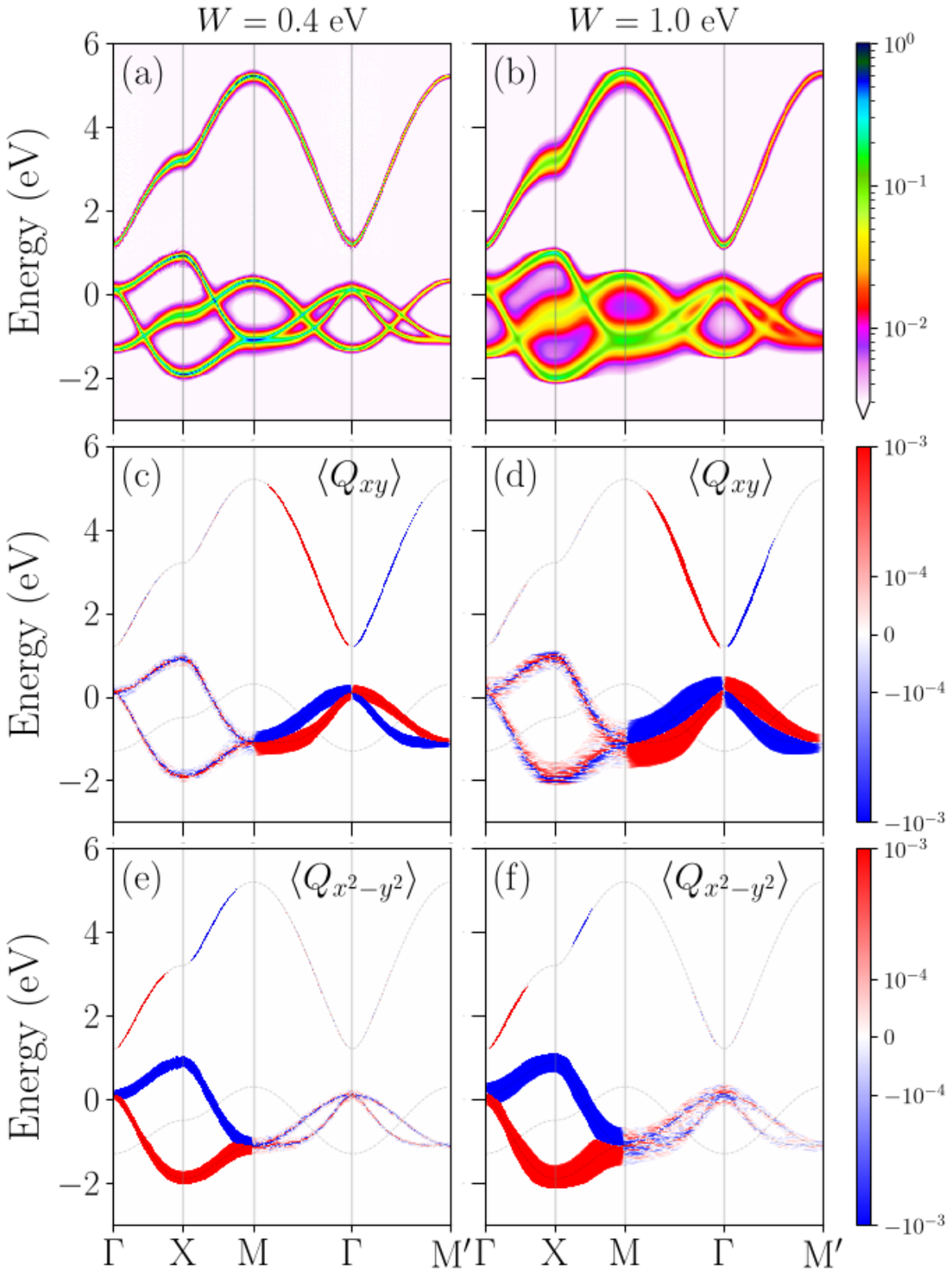}
        \caption{
        Anderson disorder. 
        Spectral and orbital-texture properties for $W=0.4$~eV (left column) and $W=1.0$~eV (right column), along the $\Gamma$-${\rm X}$-${\rm M}$-$\Gamma$-${\rm M}'$ path of Fig.~\ref{fig:appendixA}(a). 
        (a),(b) Spectral function. 
        (c),(d) $\langle Q_{xy}\rangle$ and (e),(f) $\langle Q_{x^2-y^2}\rangle$, both on a $[-10^{-3},10^{-3}]$ color scale rather than the $[-1,1]$ scale of Fig.~\ref{fig:Fig_01}: the two principal bands retain $O(1)$ texture (saturated red/blue), while the disorder-induced leakage into the nominally forbidden regions is a weak, three-order-of-magnitude-smaller correction. 
        }
        \label{fig:Fig_03}
    \end{figure}
      
    Figure~\ref{fig:Fig_03} shows the disorder-averaged spectral function together with the \(\langle Q_{xy}\rangle\) and \(\langle Q_{x^2-y^2}\rangle\) textures along the same \(\Gamma\)-X-M-\(\Gamma\)-M\(^{\prime}\) path used in Appendix~\ref{App:kite-arpes}.
    The two columns correspond to \(W=0.4\)~eV and \(W=1.0\)~eV.
    As \(W\) increases, the spectral function broadens and spectral weight spreads into regions that are forbidden in the pristine limit, reflecting the loss of exact translational symmetry.

    The quadrupolar textures show the same qualitative effect, but on a much smaller scale.
    Weak disorder-induced weight appears in the clean-limit forbidden regions: \(\langle Q_{xy}\rangle\) along \(\Gamma\)-X and \(\langle Q_{x^2-y^2}\rangle\) along M-\(\Gamma\)-M\(^{\prime}\).
    This leakage is larger for \(W=1.0\)~eV than for \(W=0.4\)~eV, consistent with stronger momentum mixing.
    However, the main quadrupolar features remain concentrated on the principal band branches and are much larger than the leakage, as indicated by the \([-10^{-3},10^{-3}]\) color scale used in Fig.~\ref{fig:Fig_03}(c)--(f).
    Thus, Anderson disorder visibly broadens the spectral function, while the dominant quadrupolar character of the main bands remains identifiable over the disorder range shown.    

    \subsection*{$V_{sp\sigma}$ Bond Disorder}

    \begin{figure}[t]
    \centering
    \includegraphics[width=1.0\linewidth]{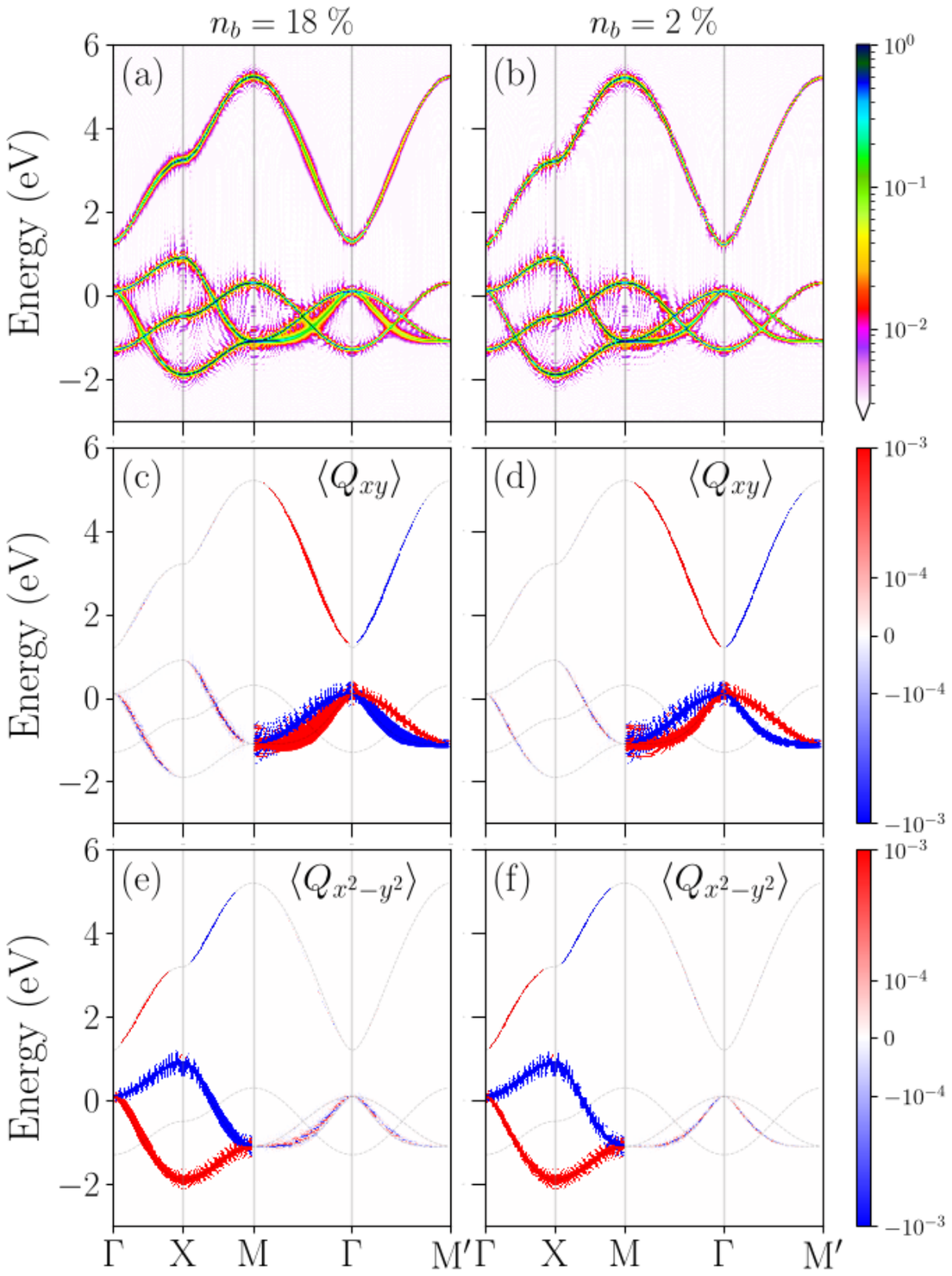}
    \caption{
    $V_{sp\sigma}$ bond disorder.
    Spectral and orbital-texture properties for $n_b=18\%$ (left column, the anomalous peak dilution fraction of Fig.~\ref{fig:Fig_06}(c)) and 
    $n_b=2\%$ (right column), 
    along the $\Gamma$-$X$-$M$-$\Gamma$-$M'$ path of Fig.~\ref{fig:appendixA}(a). 
    (a),(b) Spectral function. 
    (c),~(d)~$\langle Q_{xy}\rangle$ and (e),~(f)~$\langle Q_{x^2-y^2}\rangle$, both plotted on a $[-10^{-3}, 10^{-3}]$ color scale. 
    }
    \label{fig:Fig_07}
    \end{figure}

    Figure~\ref{fig:Fig_07} shows the corresponding spectral function and quadrupolar textures for \(n_b=18\%\), close to the transport maximum of Fig.~\ref{fig:Fig_06}(c), and for a weakly diluted case, \(n_b=2\%\), along the \(\Gamma\)-X-M-\(\Gamma\)-M\(^{\prime}\) path. 
The main effect of increasing the bond dilution is not to destroy the band structure, but to broaden and redistribute the orbital texture within the low-energy \(p_x,p_y\) manifold. 
This is particularly clear in the \(\langle Q_{xy}\rangle\) maps: at \(n_b=18\%\), the texture is more spread over the low-energy branches than at \(n_b=2\%\), reflecting the fact that this channel is directly tied to the \(s\)-\(p\) hybridization through \(C\propto V_{sp\sigma}^2\). 
Thus, bond dilution acts on the same microscopic ingredient that controls the clean \(Q_{xy}\) texture, but it does so locally and stochastically, producing spectral broadening and momentum mixing rather than a simple rigid rescaling of the clean bands.

The \(\langle Q_{x^2-y^2}\rangle\) texture is affected in a different way. 
Its coefficient \(\mathcal{B}=B+C/2\) contains a finite \(V_{sp\sigma}\)-independent contribution \(B\), so reducing the \(s\)-\(p\) hybridization does not eliminate this channel. 
Consistently, the \(\langle Q_{x^2-y^2}\rangle\) maps retain a recognizable sign structure on the low-energy branches even at \(n_b=18\%\), although the texture is broadened by disorder. 
This contrast between the directly diluted \(Q_{xy}\) channel and the partially retained \(Q_{x^2-y^2}\) channel supports the interpretation of the bond-dilution peak as a combination of clean hybridization physics and disorder-induced redistribution of spectral weight.

\section{Non-monotonic $V_{sp\sigma}$ Dependence in the Pristine Limit} 
\label{App:OHE-pristine}
\setcounter{figure}{0}
\renewcommand{\thefigure}{C\arabic{figure}}

The non-monotonic \(\sigma^{L_z}_{\rm OHE}(n_b)\) found in Sec.~\ref{SubSec:bond-disorder} suggests that bond dilution does more than simply suppress the \(s\)-\(p\) hybridization. 
To isolate the part of this behavior that is already present in the clean model, we first ask how the low-energy Berry curvature depends on a uniform change of \(V_{sp\sigma}\). 
Eq.~\eqref{eq:omega_final} shows that the projected Berry curvature contains the product \(\mathcal{A}C\mathcal{B}\), with \(C=4a^2V_{sp\sigma}^2/(\varepsilon_p-\varepsilon_s)\), \(\mathcal{A}=A+C/2\), and \(\mathcal{B}=B+C/2\). 
For the parameters used here, \(C<0\), so increasing \(V_{sp\sigma}\) increases the magnitude of \(C\) while reducing both \(\mathcal{A}\) and \(\mathcal{B}\). 
The response is therefore not expected to be monotonic in \(V_{sp\sigma}\).

This can be seen directly from the small-\(\mathbf{k}\) prefactor \(F(V_{sp\sigma})=\mathcal{A}C\mathcal{B}\). 
With \(A=0.15\)~eV, \(B=0.35\)~eV, and \(\varepsilon_p-\varepsilon_s=-3.7\)~eV, \(F\) vanishes at \(V_{sp\sigma}=0\), at \(V_{sp\sigma}=0.527\)~eV where \(\mathcal{A}=0\), and at \(V_{sp\sigma}=0.805\)~eV where \(\mathcal{B}=0\). 
The corresponding extrema occur at \(V_{sp\sigma}\simeq0.348\)~eV and \(V_{sp\sigma}\simeq0.704\)~eV. 
Thus, even in the pristine effective Hamiltonian, the orbital Hall response is controlled by a balance between the \(Q_{xy}\)-like channel associated with \(C\), the \(Q_{x^2-y^2}\)-like channel associated with \(\mathcal{B}\), and the orbital-current prefactor \(\mathcal{A}\).

\begin{figure}[h]
    \centering
    \includegraphics[width=1.00\linewidth]{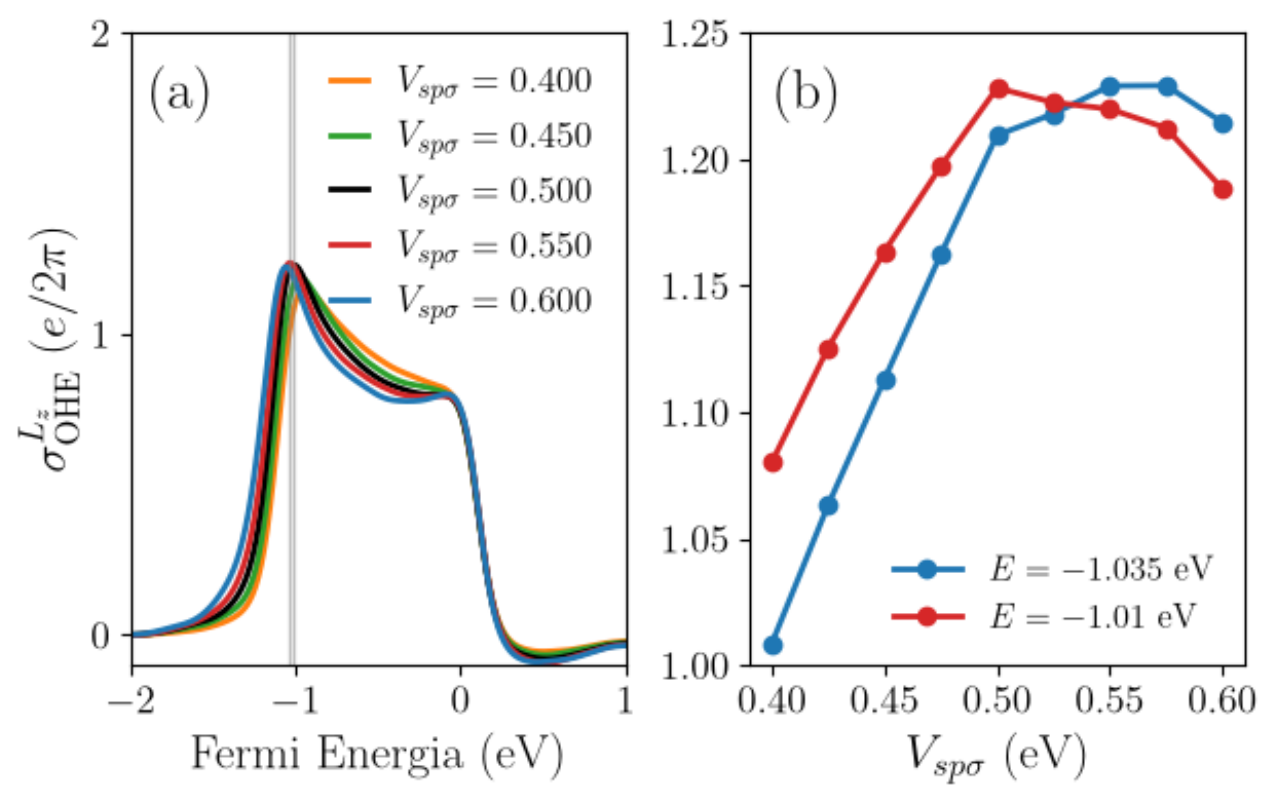}
    \caption{
    Clean full-lattice dependence on \(V_{sp\sigma}\).
    (a) \(\sigma^{L_z}_{\rm OHE}(E_F)\) for several values of \(V_{sp\sigma}\) close to the pristine value \(V_{sp\sigma}=0.50\)~eV (black). 
    Both the height and the energy position of the main orbital Hall peak near \(E_F\approx-1\)~eV depend on \(V_{sp\sigma}\).
    (b) \(\sigma^{L_z}_{\rm OHE}(V_{sp\sigma})\) at two fixed Fermi energies, \(E_F=-1.01\)~eV and \(E_F=-1.035\)~eV, illustrating that the position of the maximum is energy dependent.
    }
    \label{fig:appendixC1} 
\end{figure}

Figure~\ref{fig:appendixC1} shows that the non-monotonic dependence on \(V_{sp\sigma}\) survives beyond the small-\(\mathbf{k}\) expansion when the full lattice Hamiltonian, Eq.~\eqref{eq:real-sapce_hamiltonian}, is used without disorder. 
At \(E_F=-1.01\)~eV, \(\sigma^{L_z}_{\rm OHE}\) increases and then decreases as \(V_{sp\sigma}\) is varied from \(0.40\) to \(0.60\)~eV, with a maximum near the pristine value \(V_{sp\sigma}=0.50\)~eV. 
At the nearby energy \(E_F=-1.035\)~eV, the maximum is shifted toward \(V_{sp\sigma}\simeq0.55\)~eV. 
The full-lattice maximum is therefore not expected to coincide quantitatively with the small-\(\mathbf{k}\) estimate: the conductivity receives contributions from a finite energy window and from regions of the Brillouin zone beyond the quadratic expansion around \(\Gamma\). 
The important point is qualitative: the clean system already has a non-monotonic dependence on \(V_{sp\sigma}\).

This observation is useful for interpreting the bond-dilution result, but it should not be turned into a one-to-one mapping. 
If one makes a crude virtual-crystal estimate, a \(V_{sp\sigma}\) bond survives only when neither endpoint belongs to the diluted set, giving \(V_{sp\sigma}^{\rm eff}/V_{sp\sigma}^{0}\simeq(1-n_b)^2\). 
For \(n_b\simeq18\%\), this gives \(V_{sp\sigma}^{\rm eff}\simeq0.336\)~eV, close to the lower small-\(\mathbf{k}\) extremum at \(V_{sp\sigma}\simeq0.348\)~eV. 
This agreement is suggestive, but only at the level of the low-energy hybridization balance. 
Stochastic bond dilution is not equivalent to a uniform rescaling of \(V_{sp\sigma}\): it changes the local electronic environment around the removed bonds, redistributes spectral weight, broadens the bands, and mixes momenta. 
These disorder-specific effects can shift the maximum relative to both the small-\(\mathbf{k}\) estimate and the clean full-lattice calculation. 
The appendix therefore supports a more cautious interpretation: the bond-dilution peak appears to combine the non-monotonic clean dependence on \(s\)-\(p\) hybridization with additional spectral effects produced by random bond removal.
\section{On-site Disorder} 
\label{App:on-site_disorder}
\setcounter{figure}{0}
\renewcommand{\thefigure}{D\arabic{figure}}

As a final disorder mechanism, we consider a dilute concentration of identical, controlled on-site impurities, complementary to the continuous, all-site Anderson disorder of Sec.~\ref{SubSec:Anderson} in that here every orbital at a small, fixed fraction $n_o$ of randomly selected sites receives the same energy shift $\delta U$, $\varepsilon_i^\alpha \rightarrow \varepsilon_i^\alpha + \delta U$, $i\in\mathcal{S}$, where $\mathcal{S}$ is a random subset of sites of concentration $n_o$, and $\delta U$ is a controlled parameter rather than a random variable. 

\begin{figure}[h]
    \centering
    \includegraphics[width=1.00\linewidth]{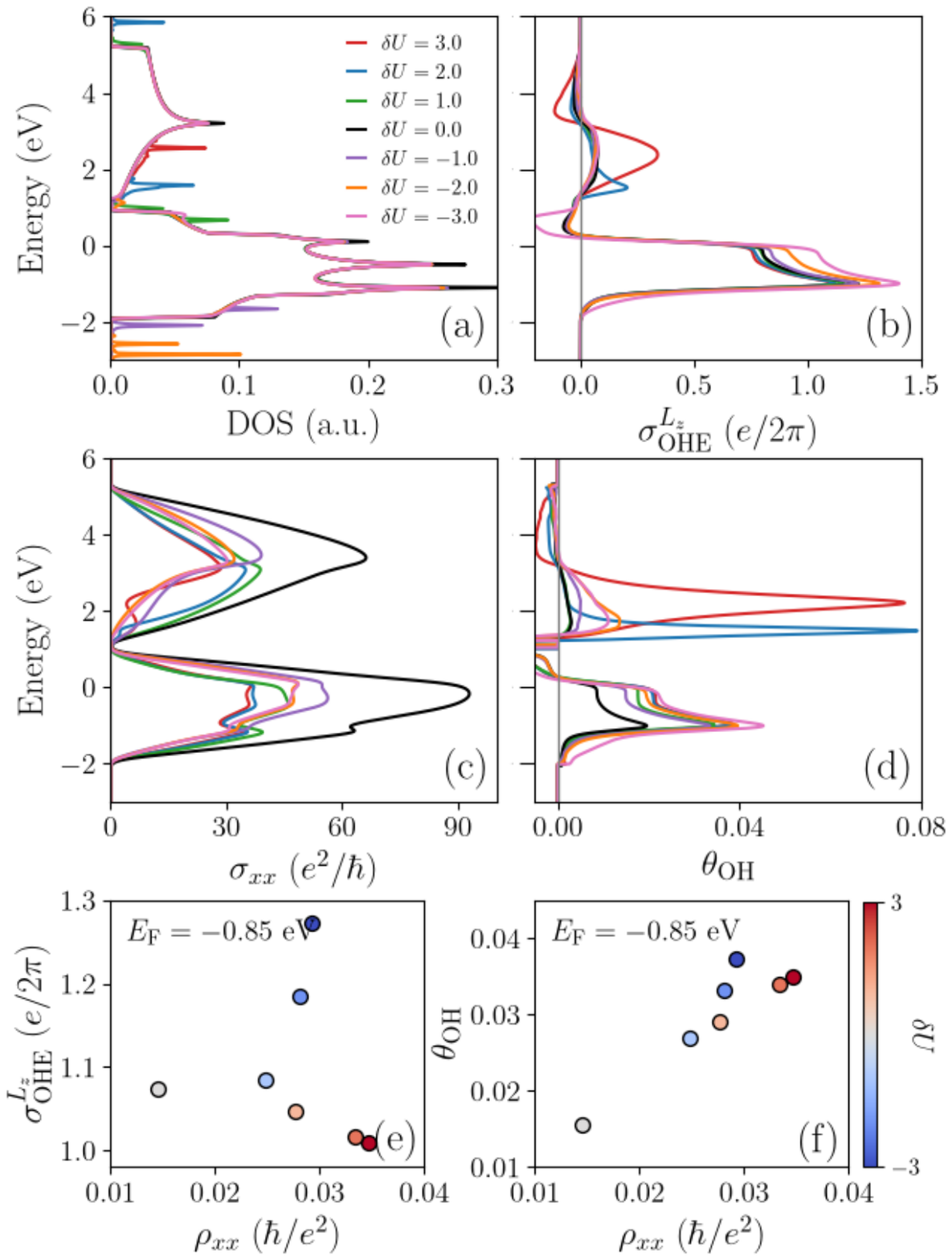}
    \caption{
    On-site disorder at fixed concentration $n_o=2\%$.
    (a) Density of states, (b) orbital Hall conductivity $\sigma^{L_z}_{\rm OHE}$, (c) longitudinal conductivity $\sigma_{xx}$, and (d) orbital Hall angle $\theta_{\rm OH}$ as a function of energy, for impurity strengths $\delta U=-3.0$ to $3.0$~eV. The black line ($\delta U=0.0$) corresponds to the pristine system.
    (e) $\sigma^{L_z}_{\rm OHE}$ and (f) $\theta_{\rm OH}$ at fixed $E_F=-0.85$~eV plotted against $\rho_{xx}$, parameterized by $\delta U$.
    }
    \label{fig:appendixD1} 
\end{figure}

Figures~\ref{fig:appendixD1}(a)-(d) show the resulting spectral and transport properties. 
Each nonzero $\delta U$ introduces sharp, resonance-like features in the density of states [panel (a)] whose energy position shifts systematically with $\delta U$, consistent with the appearance of a localized impurity level pulled away from the host bands -- a characteristic signature of a resonant impurity~\cite{Koster1954a,Koster1954b}, here realized at finite but dilute concentration.
The pristine system ($\delta U=0$, black) retains the sharpest and tallest features in $\sigma_{xx}$ [panel (c)], as expected in the absence of any impurity scattering; every nonzero $\delta U$ broadens and suppresses $\sigma_{xx}$, with $\sigma^{L_z}_{\rm OHE}$ and $\theta_{\rm OH}$ [panels (b),(d)] responding correspondingly.

Panels (e) and (f) reveal a clear asymmetry between positive and negative $\delta U$ at fixed $E_F=-0.85$~eV. 
At comparable $|\delta U|$, positive shifts ($\delta U>0$) suppress $\sigma_{xx}$ more strongly than negative shifts of the same magnitude, and correspondingly suppress $\sigma^{L_z}_{\rm OHE}$ more strongly as well. 
Thus, the two branches do not simply collapse onto a single curve parameterized by $|\delta U|$. 
This sign-dependence is the expected hallmark of resonant impurity scattering, whose efficiency depends on where the impurity level falls relative to $E_F$, rather than on the impurity strength alone; a repulsive ($\delta U>0$) and an attractive ($\delta U<0$) impurity of equal magnitude are therefore not equivalent at this Fermi energy.

\begin{figure}[h]
    \centering
    \includegraphics[width=1.00\linewidth]{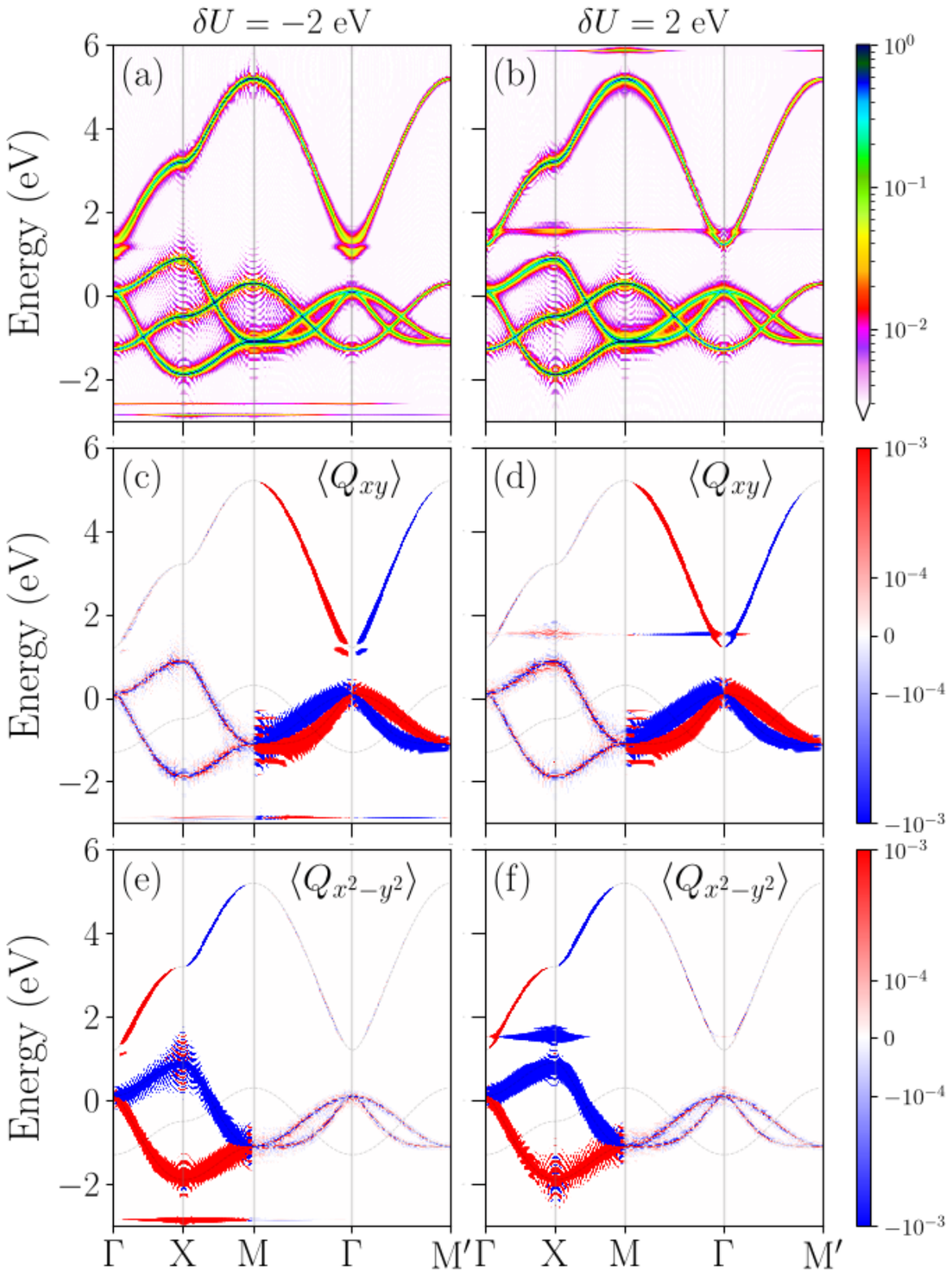}
    \caption{
    On-site disorder at fixed concentration $n_o=2\%$. 
    Spectral and orbital-texture properties for $\delta U=-2$~eV (left column) and $\delta U=+2$~eV (right column), along the $\Gamma$-$X$-$M$-$\Gamma$-$M'$ path of Fig.~\ref{fig:appendixA}(a). 
    (a),(b) Spectral function. 
    (c),(d) $\langle Q_{xy}\rangle$ and (e),(f) $\langle Q_{x^2-y^2}\rangle$.
    }
    \label{fig:appendixD2} 
\end{figure}

Figure~\ref{fig:appendixD2} shows the corresponding spectral function and orbital textures for $\delta U=-2$~eV (left column) and $\delta U=+2$~eV (right column), along the same $\Gamma$-${\rm X}$-${\rm M}$-$\Gamma$-${\rm M}'$ path used throughout this work. 
Beyond the disorder-induced leakage already seen for Anderson, vacancy, and bond disorder -- $\langle Q_{xy}\rangle$ into $\Gamma$-${\rm X}$, $\langle Q_{x^2-y^2}\rangle$ into ${\rm M}$-$\Gamma$-${\rm M}'$ -- panels (a),(b) reveal an additional, qualitatively distinct feature: narrow, nearly $\mathbf{k}$-independent bands appearing at fixed energies away from the host dispersion. 
Unlike the $\mathbf{k}$-mixing leakage common to every disorder type studied in this work, a genuinely localized or resonant impurity state disperses weakly with $\mathbf{k}$, appearing instead as a horizontal streak in the spectral function -- a characteristic signature of the dilute resonant-impurity problem~\cite{Koster1954a,Koster1954b}. 
The energy at which this feature appears shifts with the sign of $\delta U$, consistent with the impurity level's position relative to the host bands.

\section{Real-Space Orbital Textures} 
\label{App:real-space}
\setcounter{figure}{0}
\renewcommand{\thefigure}{E\arabic{figure}}

While Appendix~\ref{App:kite-arpes} resolves the quadrupole textures in momentum space, KITE's \texttt{ldos\_map()} target function additionally gives access to their real-space counterpart: the local density of states (LDOS) and the local expectation values of $Q_{xy}$, $Q_{x^2-y^2}$, and $L_z$, resolved site by site, at fixed energy $\varepsilon\approx-0.5$~eV. 
This complements the momentum-space picture by directly visualizing how each disorder mechanism perturbs the orbital texture in the region where it is actually applied.

\begin{figure}[t]
    \centering
    \includegraphics[width=1.00\linewidth]{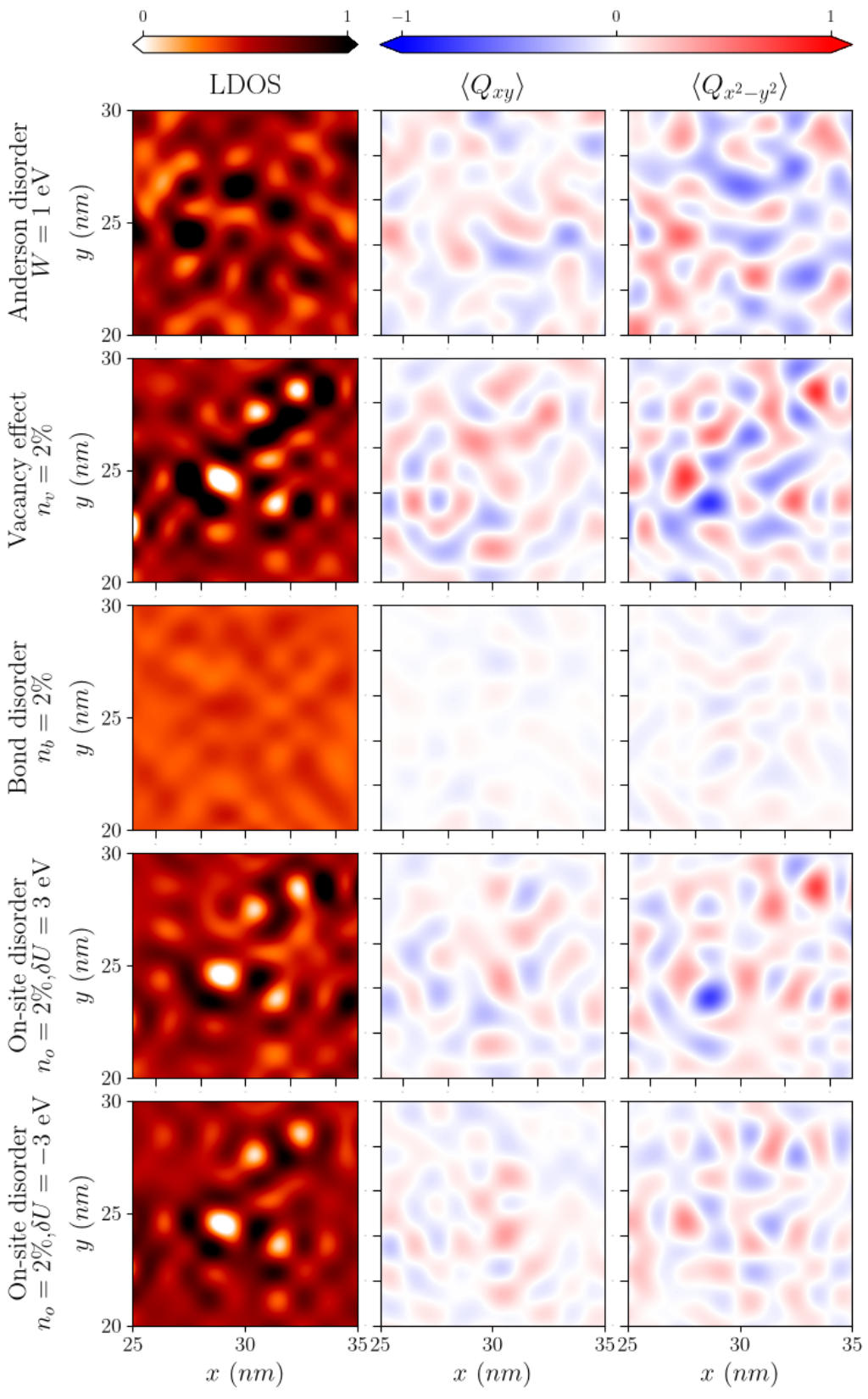}
    \caption{
    LDOS and real-space orbital textures computed via KITE over a $10\times10$~nm$^2$ window at fixed energy $\varepsilon\approx-0.5$~eV, for four disorder types at matched concentration and random seed. 
    Columns: LDOS ($p_x,p_y$ manifold only), $\langle Q_{xy}\rangle$ and $\langle Q_{x^2-y^2}\rangle$. 
    }
    \label{fig:appendixE1} 
\end{figure}

Each calculation is performed on a large system of $384\times384$~nm$^2$, but Fig.~\ref{fig:appendixE1} displays only a $10\times10$~nm$^2$ window centered on the region where disorder was introduced. 
To allow a direct, realization-by-realization comparison across disorder types at fixed concentration, the same random seed is used for vacancy and on-site disorder at $2\%$ concentration (and at $2\%$ dilution fraction for bond disorder), so that all three share the identical random site selection. 
The LDOS shown is restricted to the $p_x,p_y$ manifold (excluding $s$ and $p_z$). 

In the pristine system (not shown), the LDOS is spatially homogeneous and the local quadrupole and dipole expectation values vanish identically at every site, exactly as established in Sec.~\ref{SubSec:Microscopic_features}: the $k_x\to-k_x$, $k_y\to-k_y$ reflection symmetry forces $\langle Q_{xy}\rangle$ and $\langle Q_{x^2-y^2}\rangle$ to cancel over the full Brillouin zone at any single energy, while the reality of the Hamiltonian forces $\langle L_z\rangle=\hbar\langle\sigma_y\rangle=0$ identically for every eigenstate. 
The pristine row is therefore omitted from Fig.~\ref{fig:appendixE1}, and every pattern visible in the remaining rows is a genuine disorder-induced effect.

The dipole expectation value, $\langle L_z\rangle$, remains zero 
at every site for all four disorder types. 
This is expected: Anderson, vacancy, bond, and on-site disorder are all real perturbations of a real Hamiltonian, so every disordered eigenstate can still be chosen real, and $\langle L_z\rangle=\hbar\langle\sigma_y\rangle$ continues to vanish identically regardless of the specific disorder realization. 
By contrast, $\langle Q_{xy}\rangle$ and $\langle Q_{x^2-y^2}\rangle$ develop a markedly inhomogeneous real-space distribution once disorder is introduced, with strongly populated regions of both positive (red) and negative (blue) sign concentrated at the disordered sites themselves -- the real-space counterpart of the momentum-space leakage discussed in Secs.~\ref{SubSec:Anderson}-\ref{SubSec:bond-disorder} and Appendix~\ref{App:on-site_disorder}.

The LDOS panels reveal a distinct spatial signature for each disorder mechanism. 
Anderson disorder produces a genuinely inhomogeneous LDOS across the entire window, consistent with every site being independently perturbed. 
Vacancies instead produce sharply localized enhanced regions (high LDOS) immediately surrounding each vacant site, while the vacant site itself shows exactly zero LDOS, appearing as an isolated white spot at its center. 
Bond disorder produces a distinctive checkerboard-like pattern in the affected region, alternating between neighboring sites -- a spatial signature not seen for any of the other three disorder types. 
We verified that a single bond-diluted site reproduces the same alternation along both the axis directions away from the defect, confirming it is an intrinsic response to the $V_{sp\sigma}$ dilution itself rather than an interference effect between multiple nearby defects. 
For on-site disorder, the randomly selected impurity sites themselves show suppressed LDOS at this energy, while their immediate neighbors show enhanced LDOS, consistent with the impurity resonance discussed in Appendix~\ref{App:on-site_disorder} being centered at a different energy and locally redistributing spectral weight around it at $\varepsilon\approx-0.5$~eV.

\bibliography{ref}

\end{document}